\documentclass[11pt,a4paper]{article}
\pdfoutput=1
\usepackage{jheppub}

\usepackage{amsmath}
\usepackage{verbatim}
\usepackage{graphicx}
\usepackage{mathrsfs}
\usepackage{appendix}
\usepackage{caption}
\usepackage{float}
\usepackage{subfig}
\usepackage{mathtools}\usepackage{mathtools}

\newcommand{\gb}{\hat{\gamma}}
\newcommand{\hs}{\hat{\Sigma}}
\newcommand{\gc}{\tilde{\gamma}}

\newcommand{\be}[1]{ \begin{equation}\label{#1} }
\newcommand{\ee}{\end{equation}}
\newcommand{\bea}[1]{\begin{eqnarray}\label{#1} }
\newcommand{\eea}{\end{eqnarray}}
\newcommand{\bes}{\begin{subequations}}
\newcommand{\ees}{\end{subequations}}

\newcommand{\p}{\partial}
\renewcommand{\a}{\alpha}
\renewcommand{\b}{\beta}
\newcommand{\g}{\gamma}
\renewcommand{\t}{\tau}

\newcommand{\s}{\sigma}

\DeclarePairedDelimiterX\braket[2]{\langle}{\rangle}{#1 \delimsize\vert #2}

\newcommand{\rw}{\rightarrow}

\title{Carroll fermions in two dimensions}

\author[a]{Aritra Banerjee,} \author[b, c,d]{Sudipta Dutta,} \author[b]{and Saikat Mondal.}  \author{\\}

\affiliation[a]{Okinawa Institute of Science and Technology (OIST),
1919-1 Tancha, Onna-son, Okinawa 904-0495, JAPAN\\}
\affiliation[b]{Indian Institute of Technology Kanpur, Kalyanpur, Kanpur 208016. INDIA. \\}
\affiliation[c]{Erwin Schroedinger International Institute for Mathematics and Physics, 1090 Vienna,\\ AUSTRIA.\\}
\affiliation[d]{Institute for Theoretical Physics, TU Wien, Wiedner Hauptstrasse 8–10/136, A-1040 Vienna, AUSTRIA}
\emailAdd{aritra.banerjee@oist.jp, (dsudipta,~saikatmd)@iitk.ac.in }

\abstract{Carroll symmetry is a very powerful characteristic of generic null surfaces, as it replaces the usual Poincaré algebra with a vanishing speed of light version thereof. These symmetries have found universal applications in the physics of  null manifolds as they arise in diverse situations ranging from black hole horizons to condensed matter systems with vanishing Fermi velocities. In this work, we concentrate on fermions living on two dimensional ($2d$) null manifolds and explore the Carroll invariant structure of the associated field theories in a systematic manner. The free massless versions of these fermions are shown to exhibit $2d$ Conformal Carroll or equivalently the $3d$ Bondi-Metzner-Sachs (BMS) algebra as their symmetry. Due to the degenerate nature of the manifold, we show the presence of two distinct classes of Clifford Algebras. We also find that in two dimensions there are two distinct fermion actions. We study discrete and continuous symmetries of both theories, and quantize them using highest weight representation of the vacuum. We also discuss how the symmetries of $2d$ free fermion CFTs can be continually deformed by infinite boosts or degenerate linear transformations on coordinates, leading to the corresponding BMS invariant theory at singular points. }

\preprint{}

\begin{document}
\maketitle

\newpage

 \newpage
 \section{Introduction}
 
The road to Quantum Gravity is an uphill one, no matter how many rewards does it promise at the summit. Various approaches towards it exist, and some are more useful than others. Although most of the studies in Quantum Gravity revolves around Lorentzian (or Euclidean) structures, it makes complete sense that Non-Lorentzian symmetries would be as important in a fully consistent theory of Quantum Gravity. One of most fascinating Non-Lorentzian geometric structures of physical interest turn out to be that of a null surface. These are the manifolds on which our well known and beloved pseudo-Riemannian structures break down, forcing us to refresh our geometric point of view. These null structures could include event horizons of generic black holes, the boundary of causal diamonds, and null infinity of asymptotically flat spacetimes. But not only in the theory of Quantum Gravity, physics of null surfaces keep appearing in diverse situations, including those as unrelated as in cosmology, fluid dynamics and even spin chain systems in condensed matter physics. However, they all share one universal characteristic, they all posses some form of inherent \textit{Carrollian} symmetries that replace good old Poincaré invariance. 
\medskip 

The whimiscally named Carroll group  \cite{LevyLeblond, NDS} (after Lewis Carroll and a quote by the Red Queen in his famous book) is obtained by a seemingly bizarre Inönu-Wigner contraction of the Poincare group, where the speed of light of a system is taken to zero i.e. $c\to 0$. This group turns out to be the symmetry associated to theories living on null hypersurfaces. The associated manifolds are termed as Carrollian manifolds \cite{Henneaux:1979vn, Duval:2014lpa,Duval:2014uoa,Duval:2014uva}, as opposed to Newton-Cartan manifolds \cite{Duval:2009vt, Duval:2014uoa}, which are associated to physics in the $c\to \infty$ limit. These Carrollian manifolds have a fibre bundle structure, that keeps spatial and temporal symmetries separate. However weird the idea of zero speed of light sounds, it turns out that Carrollian limits bring out larger amounts of symmetry from relativistic parent theories, and are incredibly useful to characterize physical systems defined at highly Ultra-Relativistic (UR) regimes. In a general sense, these symmetries could arise in any theory where a characteristic notion of an effective velocity goes to zero, for example vanishing Fermi velocity in condensed matter systems should also suffice.

\medskip

The most well known application of Carroll symmetries, or more specifically the Conformal version thereof, comes in the form of  Holography in asymptotically flat spacetimes. The asymptotic symmetry group for such spacetimes at null infinity are the famous Bondi-Metzner-Sachs (BMS) groups \cite{Bondi:1962px,Sachs:1962zza}, which in dimensions three \cite{Barnich:2006av} and four are promoted to infinite dimensional structures. As mentioned before, since this is a situation where physics of null surfaces comes in handy, there should be no surprise in knowing that Conformal Carroll Algebras are isomorphic to BMS algebras in one higher dimension \cite{Bagchi:2010zz, Duval:2014uva}. This has paved way to focussed research along the directions now known as Carrollian Holography, which in the true spirit of the hologram postulates a duality between gravity in flat spacetime and a Carrollian Conformal Field Theory (CCFT) or a BMS invariant Field Theory (BMSFT) living on the null boundary \cite{Bagchi:2010zz, Bagchi:2016bcd}. This approach has found resounding success dealing with a $3d$ version thereof, where the putative boundary theory is a $2d$ CCFT \cite{Bagchi:2010zz}. See for example \cite{Bagchi:2012xr,Barnich:2012xq, Bagchi:2012cy, Bagchi:2013qva,Bagchi:2015wna,Bagchi:2014iea,Jiang:2017ecm,Barnich:2012aw, Hartong:2015usd} for a very non-exhaustive list of literature along this direction. On the other hand, the other avenue to the flatspace hologram, known as Celestial holography \cite{Strominger:2013jfa} has found incredible use in computing scattering amplitudes in the flat bulk\footnote{See the excellent reviews \cite{Strominger:2017zoo,Pasterski:2021rjz} along this direction. For more details, the keen reader is directed to the references in these reviews.}. A very recent exciting bridge between these two approaches has been established in \cite{Bagchi:2022emh}, which again uses a particular branch of a CCFT living in one less dimension\footnote{See \cite{Donnay:2022aba} for other related approaches towards bridging the two ideas.}. This makes it very clear, CCFTs are something extremely useful to study, however deeply mysterious to fathom.
\medskip

Inspired by recent advances, many authors have looked into various aspects of CCFTs in diverse dimensions. Most notably in two dimensions, where CCFT$_2$ shares another interesting duality with Galilean Conformal Field Theories (GCFT$_2$) \cite{Bagchi:2009pe}, numerous discussions have appeared in the literature. In parallel development, Carroll symmetries have also been shown to arise on null (or tensionless) string worldsheets, where they replace two copies of Virasoro as the residual symmetry algebra  \cite{Bagchi:2013bga,Bagchi:2015nca}. This has found recent and intriguing use in study of string theory near black hole horizons \cite{Bagchi:2021ban} and the explicit counting of BTZ black hole microstates \cite{Bagchi:2022iqb}. Despite the impressively long list, the intrinsic approach towards Carroll invariant theories as defined on a null hypersurface has only recently gained momentum. Carrollian conformal scalar field theories have been systematically studied from geometric point of view in \cite{deBoer:2021jej, Hao:2021urq, Saha:2022gjw}. Covariant formulations of the same has appeared in a few recent works \cite{Gupta:2020dtl, Bagchi:2022eav,Rivera-Betancour:2022lkc,Baiguera:2022lsw}. Similarly, intrinsic constructions of Carrollian gauge theories \cite{Basu:2018dub, Bagchi:2019clu,Bagchi:2019xfx,Banerjee:2020qjj, Henneaux:2021yzg}, fluid dynamics \cite{deBoer:2017ing,Ciambelli:2018wre,Ciambelli:2018xat,Ciambelli:2020eba,deBoer:2021jej} and gravity \cite{Bergshoeff:2017btm,Duval:2017els,Hartong:2015xda,Morand:2018tke,Ciambelli:2018ojf, Donnay:2019jiz, Hansen:2021fxi} has received widespread attention in recent times. 

\medskip

In this work, we will aim to fill a gap in this spectrum, by studying \textit{Carrollian fermions} in two dimensions from an intrinsic point of view. Like its scalar counterpart, the Lagrangian formulation of a Carrollian fermion was also studied first following exciting results in (Supersymmetric) null string theories \cite{Bagchi:2016yyf, Bagchi:2017cte, Bagchi:2018wsn}. In the rest of this paper, we will discuss variants of Carroll fermions arising out of first principles of symmetry. A companion paper \cite{4dcarroll} will focus on the details of Carroll fermions in general dimensions and also discuss intriguing applications to flat band physics in condensed matter systems. 

\medskip

We will make Carroll invariance our guiding principle and geometry of Carroll manifolds our beacon, thereby going ahead with introducing the representations of Carroll Clifford algebras. Due to the nature of null hypersurfaces with degenerate metrics, these Clifford elements will also be manifestly degenerate structures. We will then systematically elucidate the discrete and continuous symmetries enjoyed by a Carroll fermion in close comparison to the relativistic counterpart, at the end leading us to explicitly Carroll invariant set of Dirac actions. Depending on inequivalent representations of degenerate Clifford matrices, we recover two distinct classes of fermions, known as the homogeneous and the inhomogeneous one. We then move on to study the conformal versions of these two classes of Fermionic theories by switching off the Carroll boost invariant mass terms. We discuss the appearance of $2d$ Carroll conformal algebra (or BMS$_3$ algebra) for charges associated to relevant stress tensors.

\medskip

We then focus our attention on a rather interesting description of how Carroll objects can be found by continually deforming relativistic ones to some degenerate points. In a recent paper  \cite{Bagchi:2022nvj}, the authors have shown certain degenerate transformations (``null boosts") acting on CFT$_2$ turns it into a BMS$_3$ invariant theory at the extreme and non-invertible points. Various (anti)holomorphic CFT quantities were shown to ``flow" with the deformation induced by such a transformation, and it turns out this deformation effectively becomes a marginal current-current ($J\overline{J}$) deformation to the CFT hamiltonian, which for a particular singular value of the coupling constant makes the symmetries jump to corresponding Carrollian ones. This turns out to be a picture commensurate with the idea of adding a  
$\sqrt{T\overline{T}}$ deformations heuristically to CFT$_2$ as explored in \cite{Rodriguez:2021tcz}. Using these ideas, we will also work out how the relativistic spinors can be gradually deformed into Carroll ones as the ambient speed of light is dialed to zero. 
  \medskip
 
 The rest of this paper is organised in the following way. In section \eqref{sec2}, we introduce the notion of a Carroll Clifford algebra. Based on the two kinds of Carroll metrics, we define gamma matrices with upper indices and lower indices. Then we identify two classes of representations in two dimensions, based on whether one of the gamma matrices is purely null or not.  In section \eqref{sec3}, we talk about the Carroll Boost symmetry of fermions that replaces Lorentz symmetries in two dimensions. We also systematically derive structures associated to Dirac conjugation, Charge Conjugation and Parity transformation for Carrollian fermions. Once that is done, in section \eqref{sec4}, we write down and describe the nature of massive Dirac actions for various representations of Carroll gamma matrices. Sections \eqref{sec5}-\eqref{sec6}  contain the bulk of the main results in paper, where we talk about $2d$ massless fermions in Inhomogeneous and Homogeneous representations from a CFT point view, deriving stress tensors, algebra of charges, correlation functions and central extensions. In section \eqref{sec7}, we derive the degenerate Clifford representations and Carroll invariant actions starting from their CFT$_2$ counterpart and flowing with `boosts' to a non-invertible point. Finally, in section \eqref{sec8}, we conclude with a summary and future objectives that lay before us. A sole Appendix contains some further details on homogeneous spinors.

 \subsection*{Note added}
 While this manuscript was being prepared for communication, two papers \cite{Yu:2022bcp} and \cite{Hao:2022xhq} appeared on the arxiv, which contain related discussions on Carroll invariant fermionic fields. 
 \section{Carroll Clifford algebras}\label{sec2}
 Let us first remind ourselves quickly the basics of Carroll geometry\footnote{The reader is directed to, for example \cite{Bergshoeff:2017btm} for a more detailed introduction to this subject.}. As we mentioned in the introduction, Carroll group appears as a $c\to 0$ limit of the relativistic symmetry. The kinematical structure associated to this group ensures we can describe Carrollian manifolds in the vein of their Riemannian cousins. Intrinsically, $d$ dimensional Carrollian manifolds can be described by a fibre bundle structure with a $d-1$ dimensional base space, complemented by the temporal direction forming a one dimensional fibre.
Specifically, Flat Carroll geometries \cite{Bergshoeff:2017btm}, a particular class we are interested in, are defined using a non-degenerate metric and a no-where vanishing vector field that defines the clock-form,
 \begin{equation}\label{flat-Carroll-metric-d}
	ds^2=h_{\mu\nu}dx^{\mu}dx^{\nu} = \delta_{ij}dx^i dx^j, \qquad \tau= \frac{\partial}{\partial t}.
		\end{equation}
		These structures are then invariant under Carroll transformations, where space and time change asymmetrically under local transformations. Since the usual notion of metric is degenerate on a Carroll manifold, one cannot raise and lower indices with the metric and these two structures $(\t, h)$ step up effectively for the same role. 
		\medskip
		
In this section we will be talking about  two different kinds of Clifford algebras based on whether the Gamma matrices have a up or a down index. Since in the case of Carroll manifolds, one could define two kinds of metrics, $\t^\mu \t^\nu$ and $h_{\mu\nu}$, we could have two separate classes of Clifford algebras:
\be{defgamma}
\{\gb^\mu, \gb^\nu\} = 2\t^\mu\t^\nu,~~\{\tilde{\g}_\mu, \tilde{\g}_\nu\} = 2h_{\mu\nu}.
\ee
These the two kinds of gamma matrices (denoted by a hat and a tilde) are not a priori related to each other, and give rise to different theories living on a null manifold. For the Flat-Carroll case, we can choose our fundamental objects without a loss of generality as,
 \be{}
 \t^\mu\t^\nu = \Theta^{\mu\nu} = \text{diag}\left(1,0,0... \right),~~h_{\mu\nu} = \text{diag}\left(0,1,1... \right),
 \ee
 where the orthogonality of these two imply $\Theta^{\mu\nu}h_{\mu\nu} = 0$. As the name of this paper suggests, we will be concentrating on the two dimensional case, i.e. where both of these objects are $2\times 2$ square matrices. The definitions given in  \eqref{defgamma} will determine the representation of our Carrollian gamma matrices. In \cite{4dcarroll}, higher dimensional analogues of these representations will be described.
 
 \subsection{Representation of upper gamma matrices}
 
Consider the case of  $d=1+1$ i.e. one spacelike direction and one null direction. The Clifford algebra of the first kind, i.e. with up indices, in component form,  leads to the matrix equations: 
	\begin{eqnarray}\label{2dc}
		(\gb^0)^2 = \mathbf{I},\quad (\gb^1)^2 = \mathbf{O},\quad \{\gb^0,\gb^1\}=\mathbf{O}.
	\end{eqnarray}
This clearly a degenerate Clifford algebra, since it involves Nilpotent matrices.	
We then look for smallest non-trivial representations for these in two dimensions, in terms of $2\times 2$ matrices. The solutions of these equations $(\ref{2dc})$ can be separated into two types. The first class of solutions is called a \textit{Homogeneous} solution, deriving its name from the eponymous Super-BMS$_3$ algebra \cite{Lodato:2016alv, Bagchi:2016yyf, Bagchi:2017cte}, where $\gb^1$ is taken to be identically zero and we can choose :
	\begin{eqnarray}\label{homo11}
		\gb^0 = 
		\begin{pmatrix}
			1 & 0 \\
			0 & 1
		\end{pmatrix}
	\end{eqnarray}
 More interestingly, since the $\gb^1$ is identically zero in this case, this representation can be labelled by just the condition that $(\gb^0)^2 = \mathbf{I}$ and that would suffice for us. As a consequence any matrix with a unit square e.g. any of the hermitian Pauli matrices $(\s_1,\s_2,\s_3)$ or certain linear combinations can do the job. 
For example, consider the linear combination
\begin{align}\label{homoany}
\gb^0=a_0\bold{I}+a_i \bold{\sigma_i}, ~~a_i\in \mathbb{R}
\end{align}
Then we get,
\begin{align}
(\gb^0)^2=(a_0\bold{I}+a_i\sigma_i)(a_0\bold{I}+a_j\sigma_j) =(a_0^2+\sum_i a_i^2)\bold{I}+2a_0a_i\sigma_i.
\end{align}
In the Carroll Clifford perspective, this leads us to two distinct choices for gamma matrices for the homogeneous case: 
\begin{itemize}
\item \textbf{Case I:} \quad $a_0=1$ and $a_i=0$  \quad $\rightarrow$ \quad $\gb^0=\textbf{I}$. 
\item \textbf{Case II:} \quad  $a_0=0$ and $\sum_{i} a_i^2=1$ \quad $\rightarrow$ \quad $\gb^0=\sum_i a_i\sigma_i$.
\end{itemize}
It is easy to verify that different choices corresponding to case II are related to each other by similarity transformations and we can work with any one of the Pauli matrices by setting other coefficients to zero. However case I is a completely distinct solution as mentioned in \eqref{homo11} and can not obtained by a similarity transformation starting from case II. These substructures associated to the homogeneous representation have intriguing and subtle differences, which we will come back to later in this manuscript.
 \medskip

The other solution of the degenerate Clifford algebra is the \textit{Inhomogeneous} case, again named after the eponymous cleverly named Super-BMS$_3$ algebra \cite{Lodato:2016alv, Bagchi:2017cte}, where a class of the solutions look like:
\begin{eqnarray}\label{inho1}
\gb^1 = \begin{pmatrix}
			0 & a \\
			0 & 0
		\end{pmatrix}
\quad	\text{or} \quad
		\begin{pmatrix}
			0 & 0 \\
			a & 0 
		\end{pmatrix} ,\quad
		\gb^0
		 = 
		\begin{pmatrix}
			1 & 0 \\
			0 & -1
		\end{pmatrix}
\end{eqnarray}
Here $a$ is an arbitrary constant that can be fixed to $\pm1$ without the loss of any generality, and as we will see later, purely imaginary values are also admissible. We will mention couple of subtleties related to this value when they arise. The two classes of matrices viz. the Inhomogeneous and Homogeneous representations are not related to each other via similarity transformations due to the degenerate nature of the matrices involved.  One could now go ahead and write more varied solutions to the $(\ref{2dc})$, e.g.
	\begin{eqnarray}\label{altinhomo}
		\gb^1= 
		\begin{pmatrix}
			1 & 1 \\
			-1 & -1
		\end{pmatrix}
		,\quad
		\gb^0 = 
		\begin{pmatrix}
			0 & 1 \\
			1 & 0
		\end{pmatrix}
	\end{eqnarray}	
	but they can be related to the earlier set of matrices by a straightforward similarity transformation using the matrix:
	\be{}
	S = \begin{pmatrix}
			2 & 2 \\
			1 & -1
		\end{pmatrix}
	\ee
	So that we seemingly have just two independent class of representations for this version of the degenerate Cilifford algebra, which exhausts the choices inferred from $(\ref{2dc})$. In this manuscript, we will be mostly interested in dealing with the upper gamma representations. The implications of the lower gamma representations become more important in higher dimensions, and those will be discussed elsewhere \cite{4dcarroll}. However, for completeness, we would offer a brief glimpse into those in what follows. 
	
 \subsection{Representation of lower gamma matrices}
 
 The Clifford algebra of the second kind, with down indices, lead us to the algebraic relations: 
 \begin{eqnarray}\label{2dc2}
		(\gc_0)^2 = \mathbf{O},\quad (\gc_1)^2 = \mathbf{I},\quad \{\gc_0,\gc_1\}=\mathbf{O}.
	\end{eqnarray}
	
Interestingly, as in the previous section with up indices, this leads us to one trivial solution where $\gc_0 =\mathbf{O}$ and $\gc_1 =\mathbf{I}$  and one set of non-trivial solutions which read: 

	\begin{eqnarray}      \label{gamma}
\gc_0 = 
		\begin{pmatrix}
			0 & a \\
			0 & 0
		\end{pmatrix}
		\quad	\text{or} \quad
		\begin{pmatrix}
			0 & 0 \\
			a & 0 
		\end{pmatrix} ,\quad
		\gc_1
		 = 
		\begin{pmatrix}
			1 & 0 \\
			0 & -1
		\end{pmatrix}
	\end{eqnarray}
It is then straightforward to see that there is an implicit relationship between the representations of the upper and lower Carroll Gamma matrices in the form $\gb^1 = \gc_0$ and $\gb^0 = \gc_1$ etc. So for the lower gammas, we also have analogues of Homogeneous and Inhomogeneous set of representations. One can use these lower dimensional representations to generate higher dimensional ones via Kronecker products, and more details on this procedure can be found in \cite{4dcarroll}. Once we have fixed our representation of gamma matrices, we can go forward and define our spinor structure. 

 
 \section{Structure of Carroll spinors}\label{sec3}

Let us, for completeness, briefly remind ourselves the structure of relativistic Dirac fermions in $d=2$ (in $1+1$ signature) before we move on to the Carroll case. In these dimensions, a faithful  real representation of gamma matrices belonging to $Cl(1,1)[\mathbb{R}] $ can be given in terms of the Pauli matrices, where
\be{}
\g^0 = \s_1,~~\g^1 =-i\s_2.
\ee
The charge conjugation relations for this case read, 
\begin{eqnarray}
		\gb^{\mu^T} = -\eta \mathcal{C}\gb^{\mu}\mathcal{C}^{-1}\,, \quad \mathcal{C}^T = -\epsilon \mathcal{C}
	\end{eqnarray}	 	
for $\eta=\pm 1$ and $\epsilon=\pm 1$, which depends on spacetime dimensions, as in even dimensions both signs may work, while in odd dimensions only one would be present. But both choices would generate charge conjugation operation on the spinors, given by ${\displaystyle \psi \mapsto \psi ^{c}=\,\mathcal{C}{\overline {\psi }}^{{T}}}$. These choices would also need to satisfy $\mathcal{C}^2 = \mathbf{I}$. In even dimension it is a safe bet to choose a $\mathcal{C}$ which allows for Majorana spinors to exist.  A simple choice pertaining to these two cases is given by:
\be{}
\mathcal{C}_{+} = \s_1,~~\mathcal{C}_{-}= i\s_2.
\ee
Both of these are real under complex conjugation, and the first choice correspond to purely real Majorana spinors.  Similarly, the Dirac adjoint for the fermion is defined  as,
\begin{equation}
		\bar{\Psi} = \Psi^{\dagger}\g^0.
	\end{equation}	
	In the relativistic case with purely real spinors the charge conjugation matrix and the adjoint matrix turn out to be the same. But these matrices are representation dependent, and one needs to be careful dealing with these in degenerate Clifford case. In this section we will see how this works out for the Carroll case, by approaching the situation from first principles. 
\subsection{Defining Carrollian Dirac Adjoint}
We would go in analogy with the route one would take for Lorentzian fermions, and consider the commutators of the upper-index gamma matrices defined in  \eqref{defgamma},
	\begin{equation}
		\hs^{\mu\nu} := \frac{1}{4}[\gb^{\mu},\gb^{\nu}]
	\end{equation}		
As in the case of Lorentz generators, one could show that the commutator algebra of these $\hs$ matrices generate a Carroll algebra, which in two dimensions is nothing but Carroll boosts. It is clear from our degenerate representation of gamma matrices that only $\hs^{01}$ or $\hs^{10}$ has components that are non zero, at least in the inhomogeneous case. One can then write the action of the Carroll boost on a two component spinor $(\psi_0,\psi_1)^T$ as:
\be{bust}
\hs\psi_0 = 0,~~~\hs\psi_1 = \frac{a}{2}\psi_0.
\ee
Since there are no rotations in $2d$, these are the basic transformation laws we will be interested in. Note that we have not yet fixed the value of the parameter $a$ to $\pm 1$ or any other constant value. As a matter of fact, we will show later in the paper that the value of $a$ does not matter, and can be connected to a special automorphism of the Carroll conformal algebra.
\medskip

For the homogeneous case, since one of the gamma matrices is identically zero, all generators of the Carroll algebra vanish identically. Homogeneous fermions basically map to themselves under a Carroll transformation, i.e. invariance in this case is trivial. 
\medskip

 As in relativistic case, where we need a Lorentz invariant  bilinear form to construct the Dirac action for fermions, here also in Carroll sense we need a similar kind of bilinear form which has to be invariant under the Carrollian boosts. Following relativistic analogy we write down an ansatz for our bilinear:
	\begin{equation}
		\Psi^{\dagger}\Lambda\Psi
	\end{equation}		
where $\Lambda$ is a square, invertible matrix yet to be found by the symmetries, and dagger denotes the usual adjoint operation for matrices. We demand that Carrollian invariance requires 
	\begin{equation}\label{sd}
		\hs^{{\mu\nu}^\dagger}\Lambda + \Lambda\hs^{\mu\nu} = 0.
	\end{equation}		
We will further choose $\Lambda$ to be Hermitian matrix so that the bilinear form is real. It is easy to check that $(\ref{sd})$ is satisfied if we can find $\Lambda$ such that 
	\begin{eqnarray}\label{pm}
		&&\gb^{{\mu}\dagger} = \pm \Lambda\gb^{\mu}\Lambda^{-1} \quad\text{(either + or -)}\\
		&&\hs^{{\mu\nu}^\dagger} = -\Lambda\hs^{\mu\nu}\Lambda^{-1}.
	\end{eqnarray}
	
	The last relation is merely $(\ref{sd})$ in disguise. We will then turn to discuss the nature of the $\Lambda$ matrix next. Following the structure of the proposed invariant bilinear, this will help us to define the Dirac adjoint by
	\begin{equation}
		\bar{\Psi} = \Psi^{\dagger}\Lambda.
	\end{equation}		
So the invariant bilinear becomes $\bar{\Psi}\Psi$, which also means that we have explicitly put in the mass term in a dirac lagrangian which stays invariant (i.e. transforms as a scalar) under the global symmetries of the fermions \footnote{One can similalry show that the components of $\bar{\Psi}\gb^\mu\Psi$ change as a Carroll vector under boosts, with our choice of Dirac conjugate. }. This is a non-trivial statement for Carrollian fermions.  In what follows, we discuss the various representations of $\Lambda$ for the cases we discussed in the last section.

\subsubsection*{\underline{Adjoint for Inhomogeneous representations}:}
\subsubsection*{a) Representation with real `$a$' }	
	 In this case, we take the parameter `$a$' in the Inhomogeneous $\gb^0$ \eqref{inho1} to be real, and without the loss of generality let us put $a=1$. Here a little algebra leads us to the representation for the adjoint matrix:
	\begin{eqnarray}
		\Lambda_{R}=
			\begin{pmatrix}
				0 & i\\
				-i & 0
			\end{pmatrix}
	\end{eqnarray}		
which satisfies $\Lambda_{R} = \Lambda_{R}^{-1} = \Lambda_{R}^\dagger$ and $\Lambda_{R}^2 = 1$. 
It can be shown that 	for the representation, we have, 
	\begin{equation}
		\gb^{\mu^\dagger}=-\Lambda_{R} \gb^{\mu} \Lambda_{R}^{-1} \qquad(\mu = 0,1)
	\end{equation}
and consequently the Carrollboost  generators transform using:
	\begin{eqnarray}
		\hs^{01^\dagger}=-\Lambda_{R} \hs^{01} \Lambda_{R}^{-1}.
	\end{eqnarray}	
	
\subsubsection*{b) Representation with with imaginary `$a$' }
	We can also take `$a$' in $\gb^0$ to be purely complex$\footnote{To comply with \eqref{pm}, the parameter in $\gb^0$  has to be either real or purely imaginary.}$, and one can choose $a=i$. Here, for consistency, we can find another choice:  
	\begin{eqnarray}\label{imL}
		\Lambda_{I}=
			\begin{pmatrix}
				0 & 1\\
				1 & 0
			\end{pmatrix}
	\end{eqnarray}		
which satisfies $\Lambda_{I} = \Lambda_{I}^{-1} = \Lambda_{I}^\dagger$ and $\Lambda_{I}^2 = 1$. 
It can be shown similarly in the real representation, 
	\begin{equation}
		\gb^{\mu^\dagger}=-\Lambda_{I} \gb^{\mu} \Lambda_{I}^{-1} \quad(\mu = 0,1)
	\end{equation}
and also for the non-zero generators,
	\begin{eqnarray}
		\hs^{01^\dagger}=-\Lambda_{I} \hs^{01} \Lambda_{I}^{-1}.
	\end{eqnarray}	

\subsection*{Map between representations with real and imaginary `$a$'}
As the reader remembers, our ultimate goal is to construct well defined Dirac-like actions for our Carroll fermions.
 Consider a generic massive Dirac Lagrangian of the form 
 	\begin{eqnarray}
 		\mathcal{L} = \bar{\Psi}\gb^{\mu}\p_{\mu}\Psi - m\bar{\Psi}\Psi = \Psi^{\dagger}\Lambda\gb^{\mu}\p_{\mu}\Psi- m\Psi^{\dagger}\Lambda\Psi
 	\end{eqnarray}
The Lagrangian has to be equivalent in all kind of values of $a$ in our representation, labelled above by the subscripts R/I. So there has to exist a similarity transformation matrix $S$ such that  $S\mathcal{L}S^{-1} = \mathcal{L}$. This further implies that $S$ has to satisfy
	\begin{eqnarray}
		S\big(\Lambda_{R(I)}\gb^{\mu}_{R(I)}\p_{\mu}-m\Lambda_{R(I)}\big)S^{-1} = \Lambda_{I(R)}\gb^{\mu}_{I(R)}\p_{\mu}-m\Lambda_{I(R)}.
	\end{eqnarray}
After solving the equations one can get the explicit form of the similarity transformation matrix in $d=2$, 	
	\begin{eqnarray}\label{simsim}
		S=
			\begin{pmatrix}
				1 & 0\\
				0 & i
			\end{pmatrix}
	\end{eqnarray}	
Also it can be shown that
	\begin{eqnarray}
		&&S\gb^{\mu}_{R(I)}S^{-1} = \gb^{\mu}_{I(R)} \\
		&&S\hs^{01}_{R(I)}S^{-1} = \hs^{01}_{I(R)}.
	\end{eqnarray}
Some comments are in order at this point: looking at the structure of the transformation matrix in \eqref{simsim} and reminding ourselves that the transformation acts on the two component spinor like $\psi \to S\psi$, we deduce that the real and imaginary `$a$'  value representations are related by map of spinor components:
\be{mapho}
\psi_0 \to \psi_0,~~\psi_1\to i\psi_1,
\ee 
where $\psi_{0,1}$ can be thought of as purely real numbers. We can note from \eqref{bust} that an imaginary value of `$a$' means a purely real fermion doesn't stay real under Carroll boosts, and hence we will mostly concerned about the real case and dropping the subscripts R/I for the rest of the paper.

\subsubsection*{\underline{Adjoint for Homogeneous representations}:}
The case for defining adjoints in the homogeneous case is much less tractable. Since these fermions do not change under boosts, all bilinears of spinor components would transform as scalars. 
The reader is reminded that we had a trivial homogeneous representation with $\gb^0=\mathbf{I}, ~\gb^1 = \mathbf{0}$, while there is a more non-trivial one having $\gb^0=\mathbf{a\cdot \s}, ~\gb^1 = \mathbf{0}, ~|\bf{a}|=1.$ The structure of \eqref{pm} suggests that for the trivial case there is no discernible constraint on $\Lambda$.
\medskip

Let us now concentrate on the case of the non-trivial homogeneous representation. Although the $\gb^1$ does not give us any constraints, let us examine the $\gb^0$ case, and since we have a freedom in choice of the matrix provided by similarity transformation, let us choose $\gb^0= \s_2$. Still, the space of choice of $\Lambda$ is distinctly large due to the freedom given by $\gb^1 = \mathbf{0}$. A bit of algebra shows that choosing $\Lambda$ to be any of the Pauli matrices, or a combination thereof, will satisfy \eqref{pm} with either sign on the RHS. This makes our life very complicated when it comes to the homogeneous spinors, and we will address this at pertinent places going forward. 
\medskip

\subsection{Defining Carrollian Charge Conjugation}
Let us move to the definition of the charge conjugation operation in the Carroll case. Since we know the transposed matrices $\gb^{\mu^{T}}$ also satisfy the same Clifford algebra, there exists a charge conjugation matrix which has the property
\begin{eqnarray}\label{ccd}
		\quad\,\gb^{\mu^T} = -\mathcal{C}\gb^{\mu}\mathcal{C}^{-1},~~\mathcal{C}^\dagger \mathcal{C} = \mathbf{I}.
\end{eqnarray}		
Since we have already fixed a representation of the adjoint operators from the Carroll algebra generators, we can plug that in as the following:
\begin{eqnarray}
		\gb^{\mu^\dagger}&=&\pm\Lambda \gb^{\mu} \Lambda^{-1}\nonumber \\ 
		\implies \gb^{\mu^*}&=&\pm\Lambda \gb^{\mu^T} \Lambda^{-1} 
		=\pm (\Lambda\mathcal{C})\gb^{\mu}(\Lambda\mathcal{C})^{-1},
	\end{eqnarray}
		where note that $\mathcal{C}$ coincides with $\pm\Lambda$ matrix for real representation, since $\Lambda^2 = \mathbf{I}$. We can now discuss the charge conjugation matrices for our representations coupled with the conjugation condition ${\displaystyle \Psi \mapsto \Psi ^{c}=\,\mathcal{C}{\overline {\Psi }}^{{T}}}$.
		\medskip


In two dimensions for real values of the paremeter $a$, in the case of Inhomogeneous representation, we can see a choice for the charge conjugation matrix:
	\begin{eqnarray}
		\mathcal{C}=
			\begin{pmatrix}
				0 & -i\\
				i & 0
			\end{pmatrix}
	\end{eqnarray}		
such that it fits completely with our choice of the adjoint $\Lambda_{R}$, which we will just call $\Lambda$ in the rest of the paper. We will use this to define real spinors in the later sections. One can also explicitly check that $(\Psi^c)^c = \Psi$, indicating the reality of Inhomogeneous spinors.
	\medskip
	
		The problem with Homogeneous representations is the same as in the case of the previous section, we effectively lose constraints since one of the gamma matrices is explicitly zero. However, note there is a subtle difference between the two classes of homogeneous representations when it comes to the sign in the conjugation relation of \eqref{ccd}. One can show that the trivial and non-trivial homogeneous representations change with different signs under generic classes of $\Lambda$ and $\mathcal{C}$ matrix. Some details on this can be found in Appendix \eqref{apA}.
	\medskip

 Once the charge conjugation is put in place, a nice exercise is to look at the number current, a Noether current associated to the Dirac lagrangian, and the Conjugation thereof. This is given by $Q^\mu = i\overline\Psi\gb^\mu\Psi$ where we have $\p_\mu Q^\mu = 0$. For the Inhomogeneous case, the currents are given by:
 \bea{}
 Q^0&=& \psi_1\psi_0^*+\psi_0\psi_1^*\\
 Q^1&=&-\psi_0\psi_0^*.
\eea
Now noting the fact that $\overline{\Psi^c}= -\Psi^T\mathcal{C}$, one can show that $(Q^\mu)^c = -Q^\mu$ i.e. the currents change sign under our charge conjugation operation. Again, one of the currents is clearly zero for the homogeneous representation, rendering this discussion futile for that case. 

\subsection{A Parity transformation operator}
Continuing on our discussion of discrete symmetries of fermions, let us first try to guess the Parity operator from first principles. Our main focus again would be the Inhomogeneous representations, due to the non-trivial structure it puts forth. A convenient definition of the action of Parity reads:
\be{}
\Psi(t,x)\to \Psi' =  \mathcal{P}\Psi(t,-x),~~~\mathcal{P}^2 = \mathbf{I}.
\ee
We demand that Parity transformation operator is also a solution of the massive Dirac equation, and the $\mathcal{P}$ passes through the Dirac operator accordingly,
\be{}
(i\gb^0\p_t+i\gb^1\p_1-m)\mathcal{P}\Psi(t,-x) = \mathcal{P}(i\gb^0\p_t-i\gb^1\p_1-m)\Psi(t,-x) = 0,
\ee  
where the change in sign in the spatial derivative accounts for the mirror transformation on the spinor. A bit of algebra identifies the obvious choice for such an operator
\be{}
 \mathcal{P} = \gb^0= \begin{pmatrix}
			1 & 0 \\
			0 & -1
		\end{pmatrix}.
\ee 
Note that this is the same as relativistic choice for the Parity operator. Once this is fixed, we would like to find out how generic Dirac bilinears change under such a transformation in the Carroll case, bearing in mind that Carroll boost dictates the adjoint structure of spinors. Let us start with the Parity transformed mass term:
\bea{}
\overline{\Psi}'\Psi' = \Psi^\dagger \gb^0 \Lambda\gb^0\Psi = -\overline{\Psi}\gb^0\gb^0\Psi = -\overline{\Psi}\Psi,
\eea
that is, the usual relativistic scalar transforms with odd Parity in the Carroll case, like a pseudoscalar. This is brought about by the fact that $\Lambda$ and $\gb^0$ anticommute in this case. Similarly, for the well known Lorentzian vector under Parity, we can see 
\bea{}
\overline{\Psi}'\gb^\mu\Psi'  = \Psi^\dagger \gb^0 \Lambda\gb^\mu\gb^0\Psi = -\overline{\Psi}\gb^0\gb^\mu\gb^0\Psi &=& -\overline{\Psi}\gb^\mu\Psi ~~\mu=0,\\\nonumber 
&=& \overline{\Psi}\gb^\mu\Psi ~~~~\mu=1,
\eea
i.e. this transforms rather like a relativistic Pseudovector. This brings us to the next order of business, choosing a `third' gamma matrix for our $2d$ representation. In relativistic case, this `third' gamma would be defined as a multiplication of all other gamma matrices,
i.e.
\be{}
\gb^2 = i\gb^0\gb^1,
\ee
but in the Carroll case this would simply boil down to $\gb^2 = -i\gb^1$, which would be degenerate as $(\gb^2)^2 = 0$. To circumvent this choice for now, let us try to constrain the structure of this matrix via parity transformation properties of the bilinears otherwise knows as pseudoscalar and pseudovector in the relativistic case:
\bea{}
&&\overline{\Psi}'\gb^2\Psi' \to \overline{\Psi}\gb^2\Psi;\\
&&\overline{\Psi}'\gb^2\gb^\mu\Psi' \to \overline{\Psi}\gb^2\gb^\mu\Psi~~~\mu=0,\\
&&~~~~~~~~~~~~\to -\overline{\Psi}\gb^2\gb^\mu\Psi~~~\mu=1.
\eea
This holds up our earlier observation that Lorentzian Parity odd and even quantities are interchanged in the Carroll case. This also constraints the structure of the third gamma matrix:
\be{}
\gb^2= \begin{pmatrix}
			0 & b \\
			c & 0
		\end{pmatrix},~~b,c\in \mathbf{C}.
\ee
The extra constraint $(\gb^2)^2 = \mathbf{I}$ gives rise to $bc=1$. It means that $\gb^2 = \s_1$ is very reasonable choice, one that we will come back to later in this work. Note carefully that $\gb^2 = -i\gb^1$ could also have been a perfect choice for the above equation had it not been for the demand of Idempotence.
\medskip

The take home message of this subsection is the Parity operation seems to be exactly opposite the relativistic one for the Carroll case. Most importantly, parity even scalars in this case require the presence of a $\gb^2$, contrary to Lorentzian parlance. One can trace the roots of this revelation in noting that $\Lambda = i\gb^0\gb^2$, which being omnipresent in the Dirac bilinears, actively change their parity properties \footnote{One can also see that $\gb^2 = \s_2$ is another valid choice, for which we will get the $\Lambda_I$ mentioned in \eqref{imL}.}. This will be important going forward, as one would usually want the Dirac action to be parity even, which may not be true in the Carroll world.

 \section{Actions in different representations}\label{sec4}
Having elucidated on the representations of degenerate gamma matrices pertaining to Carroll algebra, we now move on to write down explicit two component spinor actions for the same. Owing to the degenerate nature of the gamma matrices, we can already assume that these actions will not be symmetric in space and time derivatives.

\subsection{Inhomogeneous action with real spinors}
Let us remind ourselves the rules of the game first: the gamma matrices corresponding to the inhomogeneous  representation we will be using,
\begin{align}
\gb^0 = \begin{pmatrix}
1 & 0 \\
0 & -1
\end{pmatrix} \qquad 
 \gb^1 = \begin{pmatrix}
 0 & 0 \\
 1 & 0
 \end{pmatrix}
\end{align}
As we described before, one can also check the adjoint structure
\begin{align}\label{minus}
	\gb^{{\mu}\dagger} = - \Lambda\gb^{\mu}\Lambda^{-1}
,\quad
\Lambda= \begin{pmatrix}
0 &i \\
-i &0
\end{pmatrix}
~~~~~\bar{\Psi}=\Psi^{\dagger}\Lambda.
\end{align}
We consider the action of the Dirac form with an invariant mass term:
	\begin{eqnarray}
		\mathcal{S} = \int d^2\sigma ~\bar{\Psi}(\gb^0\p_0 + \gb^1\p_1-m)\Psi
	\end{eqnarray}
	 Note that there is no explicit $i$ in the action to make the action Hermitian. This is because of manifest negative sign in (\ref{minus}).
In terms of the two components, the action takes an asymmetrical form:
\begin{align}\label{actioninh}
\int d^2\sigma~ i\left[-(\psi^*_1 \dot{\psi_0}+\psi_0^*\dot{\psi_1}-\psi^*_0 \psi^{'}_0)-m\psi^*_0\psi_1+m\psi^*_1\psi_0 \right]
\end{align}
As before, we define the charge conjugate of spinor $\Psi$ as 
	\begin{equation}\label{ccs}
		\Psi^{(c)} = -\Lambda\mathcal{C}\Psi^{*}.
	\end{equation}
 We know that under Carroll transformation a generic spinor $\Psi$ transforms as $\Psi\rw S[\Sigma]\Psi$. We can say that $(\ref{ccs})$ is a suitable definition if $\psi^{(c)}$	transforms in a similar way under Carroll transformations. One can then go ahead and show 
	\begin{eqnarray}
		\Psi^{(c)}\rw (-\Lambda\mathcal{C})S[\Sigma]^*\Psi^* = S[\Sigma](-\Lambda\mathcal{C})\Psi^* = 					S[\Sigma]\Psi^{(c)}.
	\end{eqnarray} 
Now if we impose reality condition on the spinor i.e. $\Psi^{(c)} = \Psi$, and that gives us the reality condition of spinor components 
	\begin{align}\label{realsh}
		\psi_0^{*}=\psi_{0} ,\quad \psi^*_1=\psi_1,
	\end{align}
which is precisely the charge conjugation operation corresponding to purely real components. \footnote{Note that if we had chosen imaginary $`a$' in $\gb^1$, and used the definition of adjoint accordingly, we would have found the alternate reality condition $\psi_0^{*}=\psi_{0} \quad \psi^*_1=-\psi_1$ which reflects the map mentioned in \eqref{mapho}.}
%
\medskip

The equations of motion can be obtained by varying the lagrangian \eqref{actioninh} with respect to $\psi^*_0$ and $\psi^*_1$. These turn out to be completely decoupled equations for the two spinor components:
 \begin{align} \label{eom}
 \dot{\psi_0}-m\psi_0=0 \quad \psi_0'- \dot{\psi_1}-m\psi_1 = 0
 \end{align}
Which is  the same set of equations one gets by varying the same action with respect to $\psi_0$ and $\psi_1$. A little algebra yields the solution to these equations:
\bea{}
\psi_0(t,x) &=& \psi_0(x)\exp(mt),\\ \nonumber
\psi_1(t,x) &=& \left[\psi_1(x)\exp(-mt) + \frac{1}{2m}\psi_0'(x)\exp(mt)\right].
\eea
Notice the weird decaying/growing exponential structure of these solutions. These can be traced back to the hermitian structure of the Dirac operator in this case. In the massless case, these boil down to simpler forms,
\be{}
\psi_0(t,x)=\psi_0(x),~~\psi_1(t,x)=\psi_1(x)+t\psi_0'(x).
\ee
This set of solutions will transform like a multiplet under the Carroll Conformal Algebra, something which we will come back to in a later section. 
 \subsection*{On Parity odd and even actions}
 The attentive reader may have noted that we have been talking about a form of the Dirac equation which is odd under parity due to the structure of the parity operation. We can also go ahead and write down a parity even action, which simply reads:
 \begin{eqnarray}
		\mathcal{S}_{even} = \int d^2\sigma ~\left(i\bar{\Psi}\gb^2\gb^\mu\p_\mu\Psi - m\bar{\Psi}\gb^2\Psi\right).
	\end{eqnarray}
	
Here we will be choosing  $\gb^2 = \s_1$ without any loss of generality \footnote{Note that choosing $\gb^2 = -i\gb^1$ as we discussed earlier, leads to the simplified chiral action
\be{}
\mathcal{S} = \int d^2\sigma ~\psi^*_0(i\p_0-m)\psi_0
\ee
which is Parity even and also Carroll boost invariant solely based on the fact that $\psi_0$ does not transform under Carroll. 
}. As we discussed before, this insertion of $\gb^2$ basically boils down to changing the adjoint matrix by $\Lambda =i\gb^0 $ in our case. In component form, this can be written as:
\begin{align}\label{actioninh2}
-\int d^2\sigma~ \left[i(\psi^*_0 \dot{\psi_0}+\psi_1^*\dot{\psi_1}-\psi^*_1 \psi^{'}_0)-im\psi^*_1\psi_1+im\psi^*_0\psi_0 \right]
\end{align}
An action of this form has also been used in the context of supersymmetric null strings in \cite{Bagchi:2017cte}. We can simply see that since charge conjugation matrix will have changed for this case, the reality conditions are reflected by `skewed' transformations:
\be{}
\psi^*_0 = - \psi_1,~~\psi^*_1=-\psi_0,
\ee	
upon which the above action will be the same as the odd parity action \eqref{actioninh} under a `straight' reality condition in \eqref{realsh}. So by all means, we can safely work with the odd parity action in our case, guaranteeing purely real spinors. 
\subsection{Homogeneous action}
In the case of Homogeneous spinors, we saw that the Clifford algebra allows for some freedom in $\gb^0$, since $\gb^1 = 0$ identically for this case.
To begin with, let us explicitly write the spinors as a two component complex object $\Psi = (\psi_0,\psi_1)^T$.
 A generic action for these homogeneous spinors in $2d$ will read:
\begin{eqnarray}
		\mathcal{S} = \int d^2\sigma \left[\bar{\Psi}~\gb^0\p_0 \Psi-m\bar{\Psi}\Psi\right],
	\end{eqnarray}
	i.e. only temporal derivatives will appear in the action. Both the choices we call a trivial  and a non-trivial one, does not constrain the structure of the dirac adjoint, and neither do these fermions transform under Carroll boosts. So we would fix $\gb^0 = \s_2$ and put $\Lambda$ as the Dirac adjoint operator, upon which the final action in component form reads:
	\begin{align}\label{actionhom}
\int d^2\sigma~ i(-\psi^*_0 \dot{\psi_0}-\psi_1^*\dot{\psi_1}-m\psi^*_0\psi_1+m\psi^*_1\psi_0)
\end{align}
In the massless limit this action explicitly matches up with the spinorial action obtained from tensionless string theories as discussed in \cite{Bagchi:2016yyf} \footnote{Also see for example \cite{Gamboa:1989px, Lindstrom:1990ar, Casali:2016atr}}.
In the usual way the equations of motion can be obtained by varying the lagrangian \eqref{actionhom} with respect to $\psi^*_0$ and $\psi^*_1$. These are found to be coupled first order differential equations (or de-coupled second order),
 \begin{align} \label{eom}
 \dot{\psi_0}+m\psi_1=0 \quad \dot{\psi_1}-m\psi_0 = 0
 \end{align}
The solutions to these equations are
\begin{eqnarray}
	&&\psi_{(0,1)}(t,x) = \psi_{(0,1)}(x)\exp(\pm imt).
\end{eqnarray}
Note that these solutions are explicitly oscillatory in time, as opposed to the inhomogeneous fermions. 
\subsection{Hamiltonians}
The Hamiltonian densities for our fermions can be written as:
\be{}
H = \Pi_{\psi_\a}\dot{\psi_a}- \mathcal{L},~~\a = 0,1.
\ee
For the inhomogeneous spinors as given by \eqref{actioninh}, this can be written in a component form:
\be{}
H_I=i\psi^*_0 \psi^{'}_0+im(\psi^*_0\psi_1-\psi^*_1\psi_0).
\ee
We can see for the massless case, the Hamiltonian reduces down to $H= i\psi^*_0 \psi^{'}_0$. Similarly for the homogeneous case, using the action \eqref{actionhom} leads us to:
\be{}
H_H=im(\psi^*_0\psi_1-\psi^*_1\psi_0),
\ee
i.e. just the mass term in the action\footnote{Note that the homogeneous and inhomogeneous mass terms, while having the same form, aren't actually the same since $\psi_0$ and $\psi_1$ anticommute in different ways in these two cases.}. These homogeneous fermions have the intriguing feature that the Hamiltonian vanishes when one puts the mass to zero.  
\medskip

Given these hamiltonians, we should note a few things. It has been shown \cite{Henneaux:2021yzg} that in the kinematical sense, Carroll covariance is directly implied by the following condition on the hamiltonian densities,
\be{}
[H(x),H(x') ]  = 0.
\ee
One can explicitly check that this condition is satisfied for both the densities mentioned above. In the massless homgeneous case, we have further the whole density vanishing completely. In the study of Carroll representations, one can see the Hamiltonian appears as a central term in the algebra. Due to this, there are two distinct representations of the Carroll algebra, labelled by $H=0$ and $H\neq0$ \cite{deBoer:2021jej}. One can see for our massless hamiltonians, we also have such two distinct classes. However how this classification affects the dynamics of associated fields for our spinors, will be discussed elsewhere.

\section{Inhomogeneous fermion as Carrollian CFT}\label{sec5}
In this section we discuss about the inhomogeneous fermionic CFT example, a theory which can be recovered starting from \eqref{actioninh} and putting the mass to zero. As we know, the gamma matrices used to construct this action, fall into the inhomogeneous representation of Clifford algebra in the Carrollian sense. In this section, we will show how the symmetry generators associated to this action give rise to the Carroll Conformal Algbera in two dimensions, i.e. the BMS$_3$ algebra. 
 
\subsection{Stress tensor and BMS Charges}
We start with the massless action for inhomogeneous fermions defined on a cylinder parameterised by $(\s,\t)$: 
\be{akki2}
\mathcal{S}_{inhomo}=\int d^2\sigma~ i\left[-(\psi^*_1 \dot{\psi_0}+\psi_0^*\dot{\psi_1}-\psi^*_0 \psi^{'}_0)\right].
\ee
 Under spacetime translations the generic spinor components in our actions transform as 
 \begin{equation}
 \delta \psi_{\alpha}=\xi^{\mu}\partial_{\mu}\psi_{\alpha}.
 \end{equation}
The action in \eqref{akki2} is clearly invariant under such transformations. This would allow us  to use Noether's formula to compute the stress tensor directly. 
\begin{align}
T^{\alpha}_{\beta}=\kappa\left(\frac{\partial \mathcal{L}}{\partial_{\alpha}\psi_{a}}\partial_{\beta}\psi_{a}-\delta^{\alpha}_{\beta}\mathcal{L}\right),
\end{align}
Here we have allowed an overall normalisation factor $\kappa$ in the Canonical expressions of energy momentum tensor, which we shall fix later. In components the stress tensor can be written out as
\begin{align}
T^{\tau}_{\tau}&=i\kappa\psi_0^{*} \psi_{0}' \quad T^{\tau}_{\sigma}=i\kappa(\psi_1^{*}\psi_0'+\psi_0^{*}\psi_1^{'})  \qquad \text{and} \\ \nonumber
T^{\sigma}_{\tau}&=-i\kappa\psi_0^*\dot{\psi_0} \qquad T^{\sigma}_{\sigma}=-i\kappa(\psi_1^{*}\dot{\psi_0}+\psi_0^{*}\dot{\psi_1}).
\end{align}
The stress tensor is conserved when the fields satisfy the equations of motions. Upon using the massless equations of motion ($\dot{\psi}_0 = 0$ and $\dot{\psi}_1 = \psi'_0$) the stress tensor components become
\begin{align}
T^{\tau}_{\tau}=i \kappa\psi_{0}\psi_{0}'&=-T^{\sigma}_{\sigma} \quad T^{\tau}_{\sigma}=-i\kappa(\psi_1{\psi_0}'+\psi_0{\psi_1}') \\ \nonumber
\text{and} \quad T^{\sigma}_{\tau}&=0.
\end{align}

This particular structure of the stress tensor components is typical signature of Carrollian CFTs or BMSFTs. The tracelessness as usual implies the presence of scale symmetry and the $T^{\sigma}_{\tau}=0$ equation indicates Carroll boost invariance of the theory. In two dimensions these two conditions suffice to ensure the theory is invariant under the infinite dimensional CCA$_2$ or equivalently BMS$_3$.

 \medskip

\subsection*{Algebra from charges}
\medskip

We can evaluate all the charges associated with the infinite dimensional conformal Carroll symmetries using the stress tensors as computed above. The conformal killing vector fields for $2d$ Carroll spacetime are given by
\begin{align}
\xi^0= f'(\sigma)\tau + g(\sigma) \quad \xi^1=f(\sigma)
\end{align}
where $f$ and $g$ are arbitrary functions of $\sigma $. Now the currents associated with these symmetries can be constructed as
\begin{align}
J^{\alpha}=T^{\alpha}_{\beta}\xi^{\beta}. 
\end{align}
Corresponding charges are obtained by integrating $J^0$ over a spatial slice.
\begin{align}
Q_{\xi}=&\int d\sigma ~T^{0}_{\alpha}\xi^{\alpha} \\ \nonumber
=&\int d\sigma ~[T_1 \xi^1+T_2\xi^0] \\ \nonumber
=& \int d\sigma ~[(T_1f'(\sigma)+T_2f(\sigma))\tau+T_2g(\sigma)].
\end{align}

In above equations we have renamed  $T^{\tau}_{\sigma}$ as $T_1$ and $T^{\tau}_{\tau}$ as $T_2$ and shall continue to follow this convention for the rest of the paper. Now expanding  $f$ and $g$ in terms of fourier modes, we get:
\begin{align}
f(\sigma)=\sum_{n} a_{n}e^{in\sigma} \quad g(\sigma)=\sum_{n} b_{n}e^{in\sigma}
\end{align}
In terms of these fourier modes, the charges can be written as
\begin{align}
Q=&\sum_{n}a_{n}L_{n}+b_{n}M_n \qquad \text{with} \\ \nonumber
L_{n}=&\int d\sigma~ [T_1+in\tau T_2]e^{in\sigma} \quad M_n=\int d\sigma~ T_2 e^{in\sigma}
\end{align}
For this inhomogeneous example at hand we have,
\begin{align} \label{T_1,2}
T_1= -i\kappa(\psi_1{\psi_0}'+\psi_0{\psi_1}') \quad T_2= i\kappa\psi_0'\psi_0.
\end{align}
The massless version of the equations of motion (\ref{eom}) are solved by the following mode expansions 
\begin{align} \label{mode}
\psi_0(\tau, \sigma)=\sum_{r} \beta_{r}e^{ir\sigma} \quad \psi_1(\tau,\sigma)=\sum_{r}[\gamma_r-ir\tau \beta_r]e^{ir\sigma}
\end{align}
The equal time anti-commutation relations are given by 
\begin{equation}
\{\psi(\sigma'),\Pi_{\psi}(\sigma)\}=\delta(\sigma-\sigma')
\end{equation}
In terms of the oscillators this translate to 
\begin{align} \label{anti}
\{\beta_r,\beta_s\}=0=\{\gamma_r,\gamma_s\} \quad \{\beta_r,\gamma_s\}=\delta_{r+s,0}.
\end{align}
Using these expression along with the modes expansions given in (\ref{mode}), these charges would be written as bilinears of the modes:
\begin{align}
L_n=\kappa\sum_{r}(2r+n)\beta_{-r}\gamma_{r+n} \quad \text{and} \quad M_n=\frac{\kappa}{2}\sum_{r}r\beta_{-r}\beta_{r+n}
\end{align}

At this point we fix the overall normalisation to be $\kappa=\frac{1}{2}$. We consider this input demanding the closure of the charge algebra. for 
$\kappa=\frac{1}{2}$, using (\ref{anti}) we get the standard form of the BMS$_3$ algebra, i.e.
\begin{align}
[L_n,L_m]=(n-m)L_{m+n} \quad [L_n,M_m]=(n-m)M_{n+m} \quad [M_n,M_m]=0
\end{align}
This is however the centerless part of the algebra. Later on we shall show non-zero central charges appear, when we quantize the theory.
 
 \subsection{Highest weight representation and Primary operators}

In this subsection we shall review the highest weight representation of BMS$_3$ and comment on the operators of this theory that fits into this representation. The highest weight representation for this algebra is defined by labelling the primary operators with the eigenvalues of the $L_0$ and $M_0$. As $L_0$ and $M_0$ commute, it is possible to go to a basis where they are simultaneously diagonalisable.
Action of positive integer valued $L_n$ and $M_n$ would lower the eigenvalue of $L_0$, thus to make the spectrum bounded from below,
$L_n$ and $M_n$, $\forall n\rangle0$  are chosen as annihilation operators.
\begin{align}
[L_0,O(0,0)]=\Delta O(0,0) \quad [M_0,O(0,0)]=\xi O(0,0) \\ \nonumber
[L_n,O(0,0)]=0 \quad \text{and} \quad [M_n,O(0,0)]=0 \quad \forall  n>0
\end{align}
 Then the $L_{-n}$ and $M_{-n}$ acting on $O(0,0)$ would create the decendent operators associated with those primaries.
 However very recently this construction was generelised to so called multiplet highest weight representation \cite{Hao:2021urq,  Saha:2022gjw}, where it has been shown that the previous representation is a special singlet version of more general multiplet representation. In this novel representation although the action of $L_0$ still remains diagonal but $M_0$ acts as a non-diagonal Jordan matrix.
 \begin{align}
 [L_0,\mathbf{O}(0,0)]=\Delta \mathbf{O}(0,0) \quad [M_0,\mathbf{O}(0,0)]=\boldsymbol{\xi}\mathbf{O}(0,0) \\ \nonumber
 \text{together with}  \quad
 [L_n,\mathbf{O}(0,0)]=0 \quad [M_n,\mathbf{O}(0,0)]=0 \quad \forall n>0
 \end{align}
 
 Here $\mathbf{O}\equiv(O_0,O_1,.. O_{r-1})$ is a multiplet primary of rank r and $\boldsymbol{\xi}$ is $(r \times r)$ Jordan matrix with diagnoal entries $\xi$ and sub-diagonal entries 1. For example for a rank-3 primary the matrix would look like

\begin{align}
\boldsymbol{\xi}=\begin{pmatrix}
\xi & 0 & 0 \\
1 & \xi & 0 \\
0 & 1 & \xi 
\end{pmatrix}
\end{align} 
 
For a local primary operator of rank-$r$, this representation would induce the following transformation rules under the conformal Carroll or BMS transformation 
\begin{align}
[L_n ,\mathbf{O}_a]=&-i(\partial_\sigma+inu\partial_u)\mathbf{O}_a e^{in\sigma}+n(\Delta\mathbf{I}+in\mathbb{\xi})\mathbf{O}_a e^{in\sigma} \\ \nonumber
[M_n,\mathbf{O}_a]=&\big(-i\partial_u\mathbf{O}_a+n\mathbf{\xi}\mathbf{O}_a\big)e^{in\sigma}.
\end{align}

 Returning back to our example of inhomogeneous fermionic field theory we now derive the transformation properties of the fermionic fields $\psi_0(\tau,\sigma)$ and $\psi_1(\tau,\sigma)$ by taking brackets with the BMS charges constructed previously and show that the fields transform as a multiplet of rank-2 in this novel BMS representation.
Using the mode expansions (\ref{mode}) and and the anti-commutation relations of the oscillators (\ref{anti}) the follwing relations can be easily verified
\begin{align}
[L_n,\psi_0(\tau,\sigma)]&= \sum_{r}[L_n,\beta_r]e^{-ir\sigma}  =-\frac{1}{2}\sum_{r}(2r+n)\beta_{r+n}e^{-ir\sigma}  \\ \nonumber
&=\left[-i\partial_{\sigma}\psi_0(\tau,\sigma)+\frac{1}{2}\psi_0(\tau,\sigma)\right]e^{in\sigma}
\end{align}
Similarly, one can work out,
\begin{align}
[L_n,\psi_1(\tau,\sigma)]=&\left[-i(\partial_{\sigma}+in\tau\partial_{\tau})\psi_1(\tau,\sigma)\right]e^{in\sigma} 
+n\left[\frac{1}{2}\psi_1(\tau,\sigma)+in\tau\frac{1}{2}\psi_0(\tau,\sigma)\right]e^{in\sigma}.
\end{align}
and also transformations generated by the action of the supertranslation charges are given by
\begin{align}
[M_n,\psi_0(\tau,\sigma)]=0, \quad  [M_n,\psi_1(\tau,\sigma)]&=\left(-i\partial_{\tau}\psi_1(\tau,\sigma)+\frac{n}{2}\psi_0(\tau,\sigma)\right)e^{in\sigma}.
\end{align}

Now if we take the doublet to be \{$\psi_0(\tau,\sigma),2\psi_1(\tau,\sigma)$\}, these relations clearly agree with the transformation rules of BMS rank-2 multiplet with $\Delta=\frac{1}{2}$ and $\xi=0$. Here we would like to point out that the relative factor in the primary components is 2,  and this turns out to be a consequence of our choice of gamma matrices in (\ref{gamma}). We have chosen the parameter $a=1$ previously. However in principle it is possible to choose an arbitrary number there without spoiling the Clifford algebra. Keeping a generic number would then scale the stress tensor component $T_2$. This scaling introduces an overall factor in the expression of the supertranslation charges $M_n$. Owing to its off-diagonal nature of action on the primary components this choice the arbitrary parameter would in turn show up as a relative factor in the primary fields. This transformation of the generators preserves the structure of the symmetry algebra and hence can be identified as an automorphism. 

\subsection{Correlation functions} 
Next in this section we shall compute the two point functions of these primary fields. The symmetry algebra is powerful enough to fix the two-point functions just like its relativistic counterparts. Demanding invariances under the global part of the algebra (spanned by $L_{0,\pm1},M_{0,\pm1}$) the two-point functions of these multiplet primaries have been derived in \cite{Hao:2021urq}.  For this example, we compute the two-point function of primary operators explicitly via cannonical quantisation and assuming the highest weight conditions on the vacuum.
The following condition defines the highest weight vacuum for our theory:
\begin{align} \label{highest}
L_n |0\rangle =0 \quad M_n |0\rangle=0 \quad \forall n \geq -1
\end{align}
These conditions can be met if we define the vacuum in terms of the fermionic oscillators in the following way
\begin{align}
\beta_r |0\rangle =0 \quad \gamma_r |0\rangle=0 \quad \forall r  > 0.
\end{align}
As of now we have not used any boundary conditions yet but 
whether the fermionic index $r$ runs over integral or half integrals values depends on the boundary conditions imposed on the fermionic fields. For a fermionic field theory defined on a plane, periodic boundary conditions allow only half-integer modes, whereas anti-periodic boundary conditions allow only the integer modes. However, this scenario flips when the theory is defined on a cylinder.  Here we restrict ourselves with the half integer modes only (NS sector, $r\in \mathbb{Z}+\frac{1}{2}$) because the vacuum state of the Ramond sector does not respect the desired symmetries and thus spoils the structure of the two point function  indicated by the symmetry algebra \footnote{See however \cite{Yu:2022bcp} and \cite{Hao:2022xhq} for some interesting discussions on the Ramond sector.}. Hence from now onwards we shall be assuming the indices run over half integer values only.
The oscillators in this case satisfy the following anti-commutation relations as before,
\begin{align} 
\{\beta_r,\beta_s\}=0=\{\gamma_r,\gamma_s\} \quad \{\beta_r,\gamma_s\}=\delta_{r+s,0}.
\end{align}
Now using the mode expansions in (\ref{mode}), we can write
\begin{align}
\langle\psi_0(\tau,\sigma)\psi_1(\tau',\sigma')) 
=&\sum_{r,s} \langle\beta_r\beta_s\rangle e^{-i(r\sigma+s\sigma')} =0.
\end{align}
As  $\beta$s anti commute, it can be shifted to a side where it annihilates the vacuum at the cost of a sign, hence the correlator vanishes. 
Also we can further compute:
\begin{align}
& \langle\psi_0(\tau,\sigma)\psi_1(\tau',\sigma')\rangle =\sum_{r,s} [\langle\beta_r \gamma_s\rangle-is\tau'\langle\beta_r\beta_s\rangle]e^{-i(r\sigma+s\sigma')}  \\ \nonumber
& \quad =\sum_{r>0,s<0} \langle\beta_r \gamma_s\rangle e^{-i(r\sigma+s\sigma')} =\sum_{r>0,s<0}\delta_{r+s,0}e^{-i(r\sigma+s\sigma')} =-\frac{i}{2}\csc  \left(\frac{\sigma-\sigma'}{2}\right)
\end{align}
Similarly,
\begin{align}
& \langle\psi_1(\tau,\sigma)\psi_1(\tau',\sigma')\rangle =\sum_{r,s}\langle(\gamma_r-ir\tau\beta_r)(\gamma_s-is\tau'\beta_s)\rangle e^{-i(r\sigma+s\sigma')} \\ \nonumber
& \quad =\sum_{r>0,s<0}-i[r\tau\langle\beta_r\beta_s\rangle+s\tau'\langle\gamma_r\beta_s\rangle]e^{-i(r\sigma+s\sigma')} =\frac{i}{4}(\tau-\tau')\csc\left(\frac{\sigma-\sigma'}{2}\right)\cot\left(\frac{\sigma-\sigma'}{2}\right).
\end{align}
In these previous expressions we have assumed $\tau > \tau'$, but the correlators are anti-symmetric under the exchange of the points. Hence we can call these time ordered correlation functions.

\subsubsection*{A cylinder to plane map}

It is possible to take these correlation functions on cylinder and map it to plane. For Carrollian manifolds the mapping that does the job is given by
\begin{equation}
t=i\tau e^{i\sigma} \quad x=e^{i\sigma}
\end{equation}
Here ($t,x$) denotes the coordinates on plane and $(\tau,\sigma)$ on cylinder. Now in the previous section we have shown that $\psi_0$ and $\psi_1$ transforms like a  rank-2 primary operator with $\Delta=\frac{1}{2}$ and $\xi=0$ under the conformal Carroll transformations. For this particular plane to cylinder map the transformation rules of such objects are
\begin{align}
\psi_0(t,x)=e^{-\frac{i\sigma}{2}}\psi_0(\tau,\sigma), \quad \psi_1(t,x)=e^{-\frac{i\sigma}{2}}[\psi_1(\tau,\sigma)-\frac{i}{2}\tau\psi_0(\tau,\sigma)]
\end{align}
Using these transformation rules we can map the correlators on cylinder to correlators on plane. They are given by 
\begin{align}  \label{Corplane}
\langle\psi_0(t,x)\psi_0(t',x')\rangle&=0,  \quad \langle\psi_0(t,x)\psi_1(t',x')\rangle=\frac{1}{x-x'}  \\ \nonumber
\langle\psi_1(t,x)\psi_1(t',x')\rangle&=-\frac{(t-t')}{(x-x')^2}.
\end{align}
These expressions agrees with the correlators derived from the symmetry argument.

\subsection*{Central charge}

In conformal field theories the singular terms in stress tensor OPE is entirely fixed by the symmetry algebra. A centrally extended BMS$_3$ algebra can then be shown to be equivalent to following OPEs:
\begin{align} \label{central}
T_2(t,x)T_2(t',x') &\sim 0  \nonumber\\
T_1(t,x)T_1(t',x') &\sim \left[\frac{c_L}{2(\Delta x)^4}-\frac{2c_M(\Delta t)}{(\Delta x)^5}\right]+\left[\frac{2T_1(t',x')}{(\Delta x)^2}-\frac{4\Delta t T_2(t',x')}{(\Delta x)^3}\right] \\ \nonumber
&+ \left[\frac{\partial_{x'}T_1(t',x')}{\Delta x}-\frac{\Delta t \partial_{t'} T_2(t',x')}{(\Delta x)^2}\right]  \\ \nonumber
T_1(t,x)T_2(t',x')& \sim \frac{c_M}{2(\Delta x)^4}+\frac{2T_2(t',x')}{(\Delta x)^2}+\frac{\partial_{t'}T_2(t',x')}{\Delta x}.
\end{align}
Using the (\ref{Corplane}) and (\ref{T_1,2}) it is possible to explicitly evaluate these OPEs by Wick contraction. We have already verified 
the closure of the centerless part of the charge algebra in the previous section which accounts for the less singular terms in the OPEs. What remains to be seen are the value of central charges $c_L$ and $c_M$. We evaluate these by computing the fully contracted terms in the OPEs.
\begin{align} \label{Wick}
T_1(t,x)T_1(t',x')&\sim \frac{1}{2}:(\psi_1{\psi_0}'+\psi_0{\psi_1}')(t,x)::\frac{1}{2}(\psi_1{\psi_0}'+\psi_0{\psi_1}')(t',x'):  \\ \nonumber
&\sim \frac{1}{2}\left[\partial_x\left(\frac{1}{x-x'}\right)\partial_{x'}\left(\frac{1}{x-x'}\right)-\frac{1}{x-x'}\partial_x\partial_{x'}\left(\frac{1}{x-x'}\right)\right]  \\ \nonumber
&\sim \frac{1}{2(x-x')^4}  +..
\end{align}
and similarly, 
\begin{align}
T_1(t,x)T_2(t',x')&\sim \frac{1}{2}:(\psi_1{\psi_0}'+\psi_0{\psi_1}')(t,x)::\frac{1}{2}\psi_0'\psi_0(t',x'): \\ \nonumber
&\sim 0+..
\end{align} 
 We have mapped the stress tensor components from cylinder to plane before performing the Wick contractions. Here  $\psi'_{0,1}(t,x)$ indicates derivative with respect to the spatial coordinate $x$ and
  '...' accounts for less singular terms in the OPE.  As the correlator of $\psi_0$ with itself vanishes, the fully contracted terms in $T_1(t,x)T_2(t',x')$ also vanish. This would imply $c_M=0$. This can  also be read off in the expressions of $T_1(t,x)T_1(t',x')$ as well along with $c_L=1$. 
 \section{Homogeneous fermion as Carrollian CFT}\label{sec6}
 It is clear from the last few sections that inhomogeneous fermions are the more interesting Carroll objects. But homogeneous ones have the distinct feature that any spatial derivative is absent from their actions. 
 In this section, we briefly discuss the theory of homogeneous fermions when we switch off the mass term. The action in this case becomes: 
 \be{}
S_{homo} =  \int d^2\sigma~ i(-\psi^*_0 \dot{\psi_0}-\psi_1^*\dot{\psi_1}).
 \ee
 Remember, homogeneous fermions are insensitive to Carroll boost, hence these classes of massless actions can just be thought of as two copies of the action for chiral fermions. The single versions of these chiral fermions appear from lower gamma actions, and have been studied in detail \cite{4dcarroll}. It has also been argued that theories with no spatial derivative in real space Hamiltonian and hence having a non-dispersive Hamiltonian in corresponding fourier space, should have Carrollian symmetry by construction. Hence these are no less important as Carrollian theories.
 \medskip
 
 \noindent The massless equations of motion for the spinor components read:
\be{}
\dot{\psi}_0 = 0,~\dot{\psi}_1 = 0.
\ee
The equations of motion are solved by the following mode expansions in terms of half integer modes:
\begin{align}
\psi^{(0,1)}(\tau, \sigma)=\sum_{r\in \mathbb{Z}+\frac{1}{2}} \beta_{r}^{(0,1)}e^{ir\sigma} \end{align}
It is evident from the mode expansions that all the dynamics of the system is lost and the theory essentially lives on the spatial circle. As a consequence, the action of supertranslations on the space of solution would also be trivial and the BMS$_3$ that we started off with would be truncated to one copy of Virasoro only. We can more explicitly show this by computing the charges. 
\medskip

\noindent The canonical stress tensor components for this theory would be
\begin{align}
 T^{\tau}_{\sigma}=-\frac{i}{2}\big(\psi^{*}_{0}(\sigma)\psi_{0}'(\sigma)+\psi^{*}_1(\sigma)\psi_{1}'(\sigma)\big), \quad  T^{\sigma}_{\tau}=0,
   \quad T^{\tau}_{\tau}= 0=T^{\sigma}_{\sigma} 
\end{align}
Note that $T^{\sigma}_{\tau}$ is identically zero in this case, a telltale sign of Carroll symmetry.
 Using the previous construction applied for inhomogeneous case we can immediately identify the symmetry generators
 \begin{align}
 T_1=-\frac{i}{2}\big(\psi^{*}_{0}(\sigma)\psi_{0}'(\sigma)+\psi^{*}_1(\sigma)\psi_{1}'(\sigma)\big) \quad \text{and} \quad T_2=0.
 \end{align}
Vanishing of $T_2(\tau,\sigma)$ indicates all the supertranslation charges $M_n$ drop off  causing a symmetry truncation. But we shall still have the superrotation charges  that would form one copy of Virasoro algebra. Hence this homogeneous action would effectively behave like an $1d$ CFT. Because of this symmetry truncation, the Carroll boost acts on the field components $\psi_{0,1}(\tau,\sigma)$ trivially, as we have seen before. As the spin part in the field transformation vanishes, these objects effectively behaves like a scalar, although they are still grassmannian in nature.
The Viarasoro charges would just  be the fourier modes of the sterss tensor component $T_1(\sigma)$
\begin{align}
L_n&=-\frac{i}{2}\int d\sigma \big(\psi_0(\sigma)\psi'_0(\sigma))+\psi_1(\sigma)\psi'_1(\sigma)\big)  e^{in\sigma}  \\ \nonumber
&=\frac{1}{2}\sum_{r}(2r+n)\left[\beta^{0}_{-r}\beta^{0}_{r+n}+\beta^{1}_{-r}\beta^1_{r+n}\right].
\end{align}
In a previous section we have discussed that the choice of the Charge conjugation matrix remains arbitrary owing to the fact one of the gamma matrix is identically equal to zero. Here then we have the freedom to work with purely real spinor components, i.e $\psi^{*}_{0,1}(\sigma)=\psi_{0,1}(\sigma)$. Cannonical commutation relations after implementing these reality condition would become
 \begin{align}
 \{\psi_0(\sigma),\psi_0(\sigma')\}=\{\psi_1(\sigma),\psi_1(\sigma')\}=\delta (\sigma-\sigma')
 \end{align}
In terms of the oscillators this breaks down to 
\begin{align} \label{oscillator}
\{\beta^0_r,\beta^0_s\}=\delta_{r+s,0} \quad \{\beta^1_r,\beta^1_s \}=\delta_{r+s,0}.
\end{align}
This would straightforwardly imply the algebra:
\begin{equation}
[L_n,L_m]=(n-m)L_{n+m},
\end{equation}
i.e. just a single centerless Virasoro, as predicted.
Now we go ahead and define a Highest-Weight vacuum in terms of these oscillators 
\begin{align}
\beta^0_r |0\rangle =0 \quad \beta^1_r  |0\rangle=0 \quad \forall r  > 0
\end{align}
Using the anti-commutation relations of the oscillator modes one can easily find out the correlation functions on the cylinder
\begin{eqnarray}
&&\langle\psi_0(\tau,\sigma)\psi_0(\tau',\sigma')) 
=\sum_{r,s} \langle\beta^0_r\beta^0_s\rangle e^{-i(r\sigma+s\sigma')} = \frac{i}{2}\csc\left(\frac{\sigma-\sigma'}{2}\right) \\
&&\langle\psi_1(\tau,\sigma)\psi_1(\tau',\sigma')) 
=\sum_{r,s} \langle\beta^1_r\beta^1_s\rangle e^{-i(r\sigma+s\sigma')} = \frac{i}{2}\csc\left(\frac{\sigma-\sigma'}{2}\right) \\
&&\langle\psi_0(\tau,\sigma)\psi_1(\tau',\sigma')) = 0
\end{eqnarray}
Again, the central extensions of the algebra remains to be found. One can evaluate these charges by calculating the fully contracted terms in the OPEs as in \eqref{central}. This requires one to map cylinder correlators to the plane as before, and eventually lead to:
\begin{align} 
T_1(t,x)T_1(t',x')&\sim \frac{1}{2}:(\psi_0{\psi_0}'+\psi_1{\psi_1}')(t,x)::\frac{1}{2}(\psi_0{\psi_0}'+\psi_1{\psi_1}')(t',x'):  \\ \nonumber
&\sim \frac{1}{2}[\partial_x\left(\frac{1}{x-x'}\right)\partial_{x'}\left(\frac{1}{x-x'}\right)-\frac{1}{x-x'}\partial_x\partial_{x'}\left(\frac{1}{x-x'}\right)]  \\ \nonumber
&\sim \frac{1}{2(x-x')^4}  +..\\
T_1(t,x)T_2(t',x')&\sim 0 \\ \nonumber
\end{align}
One can easily read off from here the values $c_L=1$ and $c_M=0$. \\

It is interesting to note here that for both classes of fermionic theories, we have found the same central charges, i.e. $c_L=1$ and $c_M=0$. But these are very different theories, even in terms of underlying symmetries. In the inhomogeneous case, the BMS$_3$ is intact while in the case we have discussed above, the homogeneous one, BMS$_3$ reduces to its Virasoro sub-algebra with the explicit vanishing of the supertranslation charges. So it is clear that just the central term does not determine the representation theory and the chiral truncation is not immediate with $c_M=0$. This $c_M=0$ may or may not lead to a chiral truncation and this depends on the multiplet structure of the underlying Carrollian CFT. 

\medskip 

We note another curious feature before we conclude this section. When we consider supersymmetry, the homogeneous version of the Super BMS$_3$ \cite{Bagchi:2016yyf}, which arises from putting together the Carroll boson and the homogeneous fermion as we have discussed above, is given by:
\bea{sbmsh}
&& [L_n, L_m] = (n-m) L_{n+m} + \frac{c_L}{12} \, (n^3 -n) \delta_{n+m,0}, \nonumber\\
&& [L_n, M_m] = (n-m) M_{n+m} + \frac{c_M}{12} \, (n^3 -n) \delta_{n+m,0}, \\
&& [L_n, Q^\a_r] = \Big(\frac{n}{2} - r\Big) Q^\a_{n+r}, \quad \{Q^\a_r, Q^\b_s \} = \delta^{\a\b} \left[M_{r+s} + \frac{c_M}{6} \Big(r^2 - \frac{1}{4}\Big)  \delta_{r+s,0} \right]. \nonumber
\eea

Here $Q^{\a}, ~ \a=(0,1)$ are the supercharges and all other brackets are zero.
 Notice that this does not contain a Super-Virasoro subalgebra and hence if one attempts a truncation by turning off the supertranslations, one needs to set the conformal supercharges to zero. The truncation then leads to turning off supersymmetry as well. The inhomogenous Super BMS on the other hand, obtained by putting together the inhomogeneous fermions and Carroll bosons, as shown for the tensionless inhomogeneous superstring \cite{Bagchi:2017cte}, is given by:
 \bea{sbmsi}
&& [L_n, L_m] = (n-m) L_{n+m} + \frac{c_L}{12} (n^3 -n) \delta_{n+m,0}, \nonumber\\
&& [L_n, M_m] = (n-m) M_{n+m} + \frac{c_M}{12} (n^3 -n) \delta_{n+m,0}, \\
&& [L_n, G_r] = \Big(\frac{n}{2} -r\Big) G_{n+r}, \ [L_n, H_r] = \Big(\frac{n}{2} -r\Big) H_{n+r}, \ [M_n, G_r] = \Big(\frac{n}{2} -r\Big) H_{n+r}, \nonumber\\
&& \{ G_r, G_s \} = 2 L_{r+s} + \frac{c_L}{3} \Big(r^2 - \frac{1}{4}\Big)   \delta_{r+s,0}, \ \{ G_r, H_s \} = 2 M_{r+s} + \frac{c_M}{3} \Big(r^2 - \frac{1}{4}\Big)   \delta_{r+s,0}.\nonumber
\eea
 
 Again $G,H$ are the supercharges in this case. 
  This {\em does have} a super-Virasoro subalgebra and one can truncate the theory down to this super-chiral sector \cite{Bagchi:2018ryy}. So, evidently supersymmetrization changes the notion of a chiral truncation in these two cases.

\section{``Boosting'' to Carroll Fermions}\label{sec7}
In the last few sections, we have been talking about looking at Carroll fermions from an intrinsic point of view, i.e. using Carroll structures to define spinors attached to the covering space of a Carrollian manifold. In this section we will take an alternative viewpoint, i.e. to define `flowed' representations for fermions as they gradually go from relativistic to the ultrarelativistic $(c\to 0)$ regime. The heart of this procedure lies in using linear transformations (or special ``Boosts'') that effectively change the characteristic speed of light associated to the physical system. Although finite transformations would not change the nature of physics for such systems, one could show \cite{Bagchi:2022nvj} that for certain singular limits of such transformations, one can interpolate from a relativistic CFT regime to the ultrarelativistic BMS counterpart thereof\footnote{See also \cite{Rodriguez:2021tcz} for a related idea for deforming Virasoro symmetries into BMS one with a term added ad-hoc to the Hamiltonian.}.  In this sense, these singular transformations are equivalent to the Inönü-Wigner contraction procedures, but gives one a better handle on consistently defining the theory away from both the extremal points. 
\medskip

For the $2d$ scalar field case in \cite{Bagchi:2022nvj}, this procedure was materialized by transforming the holomorphic coordinates using a $SL(2)$ transformation (which would be a Lorentz boost had it been used on physical coordinates, say $\t,\s$ on a cylinder) and showing that BMS symmetries arise when the transformation matrix degenerates in certain limits of the ``boost'' parameter. Basically this transformation continually dials the speed of light until the lightcone closes unto itself. Unless the actual closure happens, one can always undo this transformation and a CFT remains one. We would be taking a slightly different path here, since not all fermion representations are defined on a chiral basis. But generically we'll still be looking at quantities as the speed of light is being dialed upto the singular point where Carroll structures, already discussed in previous sections, effectively emerge.
\medskip

\subsection{``Flowed'' representations of the Clifford algebra}

Let us start with the explicit realisation of the relativistic Clifford algebra again,
\be{}
\{\g^\mu, \g^\nu\}  = -2c^2\eta^{\mu\nu} =2 ~\text{diag}(1,-c^2).
\ee
Here this scaling with $c^2$ is important to keep in the sense that we would be taking the $c\to 0$ expansion of the Minkowski metric, which reads:
\be{}
\eta^{\mu\nu} =- \frac{1}{c^2}\t^\mu\t^\nu+h^{\mu\nu}+...
\ee
which clearly demands that the leading order representation in this expansion will come from the  $\t^\mu\t^\nu$ term, as we have already witnessed before in \eqref{defgamma}. Now we want to find out a representation of the gamma matrices which remain valid as $c$ dials down to zero. One such flowed representation will read:
\be{}
\{\g^\mu_F, \g^\nu_F\}  = 2 ~\text{diag}(1,-\a^2).
\ee
where $\a$ is some dimensionless quantity that interpolates between 1 and 0 corresponding to a relativistic and Carroll theory respectively. This means the gamma matrices have to satisfy the following relations,
\be{relgam}
(\g^0_F)^2 = \mathbf{I},\quad (\g^1_F)^2 = -\a^2\mathbf{I},\quad \{\g^0_F,\g^1_F\}=\mathbf{O}.
\ee

A simple way to solve this set of equations is to fix $\g^0=\g^0_F$, and thereby try to solve for the components of $\g_1^F$ using the ansatz of expanding it with Pauli matrices and Unity i.e. $\g_1^F = \bf{\lambda}\cdot\bf{\s}+\b\mathbf{I}$. So we could say, fixing the $\g^0$ fixes our flowed representation. let us, for instructive purposes, start with the Majorana representation in two dimensions, instead of the usual Majorana-Weyl. In this representation the relativistic gamma matrices are given by:
\be{}
\g^0 = \s_3,~~\g^1 =i\s_2.
\ee
Solving the conditions in \eqref{relgam} consistently, we arrive at a solution:
\be{fflow}
\left[\g^1_F(\a)\right]^2 = (\lambda_1^2+\lambda_2^2)\mathbf{I}, ~~~\lambda_3, \beta =0.
\ee
Since $\s_2$ imaginary, we have $\lambda_2^2<0$ here. This equation could have two classes of consistent solutions. In the first case we can compare it to \eqref{relgam}, and for finite values of the speed of light, we see both $\lambda_1$ and $\lambda_2$ can just be proportional to $\a$, making sure the right hand side of \eqref{fflow} goes to zero in the extreme limit of $\a=0$, making all entries of the matrix zero. Clearly at the extreme limit these set of matrices reproduce the homogeneous representation since $\g^1$ is identically zero for this case, and as we have discussed any $\g^0$ with unit square alongwith a zero $\g^1$ will boil down to the homogeneous representation (see \eqref{homoany}).
 \medskip
 
 The other class of solution is much more interesting,  the parameters are not proportional to the speed of light parameter and a simple solution indicates :
 \be{lamfag}
 \lambda_1= \frac{1-\a^2}{2},~~\lambda_2= i\frac{1+\a^2}{2}.
 \ee
 Combining everything, this leads to a flowed representation of the gamma matrices along the worldline where the speed of light varies along the lightcone:
 \bea{}
 \g^0_F=\g^0, \quad \g^1_F(\a) = \lambda_1\s_1+\lambda_2\s_2= \frac{1+\a^2}{2}\g^1+\frac{1-\a^2}{2}\g^0\g^1
 \eea
 where we assume all gamma matrices are purely real. It is now very clear that 
 \be{}
 \g^1_F(\a)=\begin{pmatrix}
			0 & 1 \\
			-\a^2 & 0 
		\end{pmatrix}
 \ee
 which makes perfect sense since this representation gives one the Majorana $\g^1$ for $\a=1$, but flows to the Carroll degenerate $\hat{\g}^1$ for $\a=0$. \footnote{Note that one could have deduced $\g^1_F(\a)=\begin{pmatrix}
			0 & \a^2 \\
			-1 & 0 
		\end{pmatrix}$  using $\lambda_1 =-\frac{1-\a^2}{2} $, which is another consistent choice. This justifies the other choice of $\hat{\g}^1$ as mentioned in \eqref{inho1}. }
 \medskip
 
We of course could have chosen any non-degenerate representation in two dimensions to start with, like the Majorana-Weyl representation with  $\g^0 = \s_1,~~\g^1 =-i\s_2$. As before, taking the ansatz $\g^0=\g^0_F$ and $\g_1^F = \bf{\lambda}\cdot\bf{\s}+\b\mathbf{I}$ we could find in this case, 
\be{ffflow}
\left[\g^1_F(\a)\right]^2 = (\lambda_2^2+\lambda_3^2)\mathbf{I}, ~~~\lambda_1, \beta =0.
\ee
So as before, we can still have a homogeneous representation by choosing $\lambda_{2,3}$ proportional to $\a$, or we could choose analogues of the solutions in \eqref{lamfag} to write the flowed representations:
\bea{}
 \g^0_F=\g^0, \quad \g^1_F(\a) = \lambda_2\s_2+\lambda_3\s_3=\begin{pmatrix}
			\frac{1-\a^2}{2} & \frac{1+\a^2}{2} \\
			-\frac{1+\a^2}{2} & -\frac{1-\a^2}{2} 
		\end{pmatrix}
 \eea
 As we can see, this gives the Majorana-Weyl choice of $\g^1$ for $\a=1$, but for $\a=0$ this remarkably boils down to the choice of inhomogeneous representations we mentioned in \eqref{altinhomo}. As a consequence the flowed inhomogeneous versions of the Majorana-Weyl and Majorana representations are related to each other via similarity transformations. It is of course better to reiterate for completeness, these are in no way related to the homogeneous class of representations via any transformations, since  $\g^1=\mathbf{0}$ in that case.
 
 \subsection{Degenerate transformations on the spinors}
 
 From last section, we saw how the Clifford elements change under a change of the speed of light. Now the point is, we explicitly want transformations of gamma matrices that read:
  \bea{}
\g^0 \to S^{-1}(\a) \g^0 S(\a), \quad \g^1\to S^{-1}(\a) \g^1 S(\a).
 \eea
 So that these transformations equivalently act as a boost on the spinors 
\be{}
\psi \to S(\a)\psi.
\ee 
And this is a transformation that remains smoothly valid over the range $1\geq\a\geq0$.
Note that, this is in general how a Lorentz boost acts on a spinor, however in this case we demand a worldline where
the $\a$ keeps changing. As mentioned earlier, in two dimensional conformal systems, one can always re-orient the system to restore 
the Lorentz frame, unless we go to a singular limit of such `boosts'. 
We then demand that the Dirac equation remains invariant under such a finite transformation, i.e.
\be{}
\bar{\psi}  \tilde{\gamma}^\mu(\a) \p_\mu \psi = \bar{\psi}(\a) \gamma^\mu \p_\mu(\a) \psi(\a) \label{equ}.
\ee{}
To show how this can be packaged into the degenerate boost paradigm, let us start with the relativistic action with Majorana representation:
\begin{align}\label{actionMajo}
S_F^{(M)}= \int d^2\sigma ~i(\psi_1 \dot{\psi_0}+\psi_0\dot{\psi_1}-\psi_0 \psi^{'}_0-\psi_1 \psi^{'}_1).
\end{align}
We need to find a transformation that boosts this action into our inhomogeneous action.  Let us first write down this lagrangian in the suggestive form
\be{}
\mathcal{L}_F^{(M)} =\frac{i}{2}\left( \psi_{\bar{z}}\partial_z\psi_{\bar{z}}-\psi_z\partial_{\bar{z}} \psi_z\right)
\ee
where on a cylinder $z,\bar{z}=\t\pm\s$ and $\psi_{z,\bar{z}} = \psi_0 \pm \psi_1$.
Now following \cite{Bagchi:2022nvj} we define the transformations that mix holomorphic and antiholomorphic coordinates:
 \be{}
\label{Symm:z_LR}
	\begin{pmatrix}
	z_{\rm L} \\ z_{\rm R}
	\end{pmatrix}
	=
	\begin{pmatrix}
	\cosh\phi & -\sinh\phi \\ -\sinh\phi & \cosh\phi
	\end{pmatrix}
	\begin{pmatrix}
	z \\ \overline{z}
	\end{pmatrix}
	\qquad
	\quad
	\ee
Let us use a parameterization where $\cosh\phi = \frac{1}{\sqrt{1-\lambda^2}}$ and $\sinh\phi = \frac{\lambda}{\sqrt{1-\lambda^2}}$ and $\lambda \in \mathbf{R}$. This is a perfectly well defined SL($2,R$) transformation when  $\lambda\neq\pm1$, and basically acts as a Lorentz boost on the holomorphic coordinates. Similarly, we can define a transformation for the spinors in this basis:
 \be{}
\label{Symm:psi_LR}
	\begin{pmatrix}
	\psi_{\rm L} \\ \psi_{\rm R}
	\end{pmatrix}
	=
	\begin{pmatrix}
	\cosh\phi & \sinh\phi \\ \sinh\phi & \cosh\phi
	\end{pmatrix}
	\begin{pmatrix}
	\psi_z \\ \psi_{\bar{z}}
	\end{pmatrix}
	\qquad
	\quad
	\ee
Note that at the level of the original spinors $\psi_{0,1}$ this could have been written as a direct boost transformation with
\be{} 
S(\a) = \begin{pmatrix}\frac{1}{\sqrt{\a}} & 0 \\ 0 & \sqrt{\a} \end{pmatrix},~~\a= e^{-2\phi}
\ee 
While for the physical coordinates $(\t,\s)$ this is precisely an inhomogeneous relative scaling $\t \to e^{-\phi} \t$ and $\s \to e^{\phi}\s$.
And indeed at the degenerate point $\lambda \to 1$ (i.e. $\a \to 0$), the action boils down to \footnote{Note that using $S^{-1}(\a)$ as the transformation matrix, i.e. taking the -ve sign of $\sinh\phi$ in \eqref{Symm:psi_LR}, would lead us to
\begin{equation}\label{actionMajo3}
\tilde{S}_F= \int d^2\sigma~i (\psi_1 \dot{\psi_0}+\psi_0\dot{\psi_1}+\psi_1 \psi^{'}_1),
\end{equation}
which corresponds to taking the other choice of $\hat{\g}^1$ as mentioned in \eqref{inho1}. },
\begin{align}\label{actionMajo2}
\tilde{S}_F^{(M)}= \int d^2\sigma~i (\psi_1 \dot{\psi_0}+\psi_0\dot{\psi_1}-\psi_0 \psi^{'}_0).
\end{align}
where we have absorbed the extra overall scaling by $\tilde{S}=e^\phi S$, which could also be absorbed into a coupling constant. The above equation is exactly the inhomogeneous action we mentioned in \eqref{akki2}.
\medskip

We could also do a similar procedure starting with the Majorana-Weyl representation of relativistic fermions. Recall the lagrangian written in the Chiral basis reads:
\be{}
\mathcal{L}_F^{(MW)} = \frac{i}{2}\left(\psi_{\bar{z}}\partial_z\psi_{\bar{z}}+\psi_z\partial_{\bar{z}} \psi_z\right).
\ee
A similar scaling of the spinor components and an extra overall scaling by $e^{3\phi}$ leads one to
\be{}
\mathcal{\tilde{L}}_F^{(MW)}  = i\psi_0\dot{\psi}_0.
\ee
This is exactly the class of actions discussed in \cite{4dcarroll}, albeit in higher dimensions, and are Carroll invariant trivially. So one could heuristically call these a cousin of Weyl actions in the relativistic case, although no such notion exists when we go to the Carrollian regime of the theory. 
\medskip

Note that the difference between the overall scaling factors for these two classes of flowed actions represent that they appear at different orders when one expands relativistic spinors actions in powers of $1/c$. This may remind one of the `Electric' and `Magnetic' branches of the Carroll actions that appear in leading and next-to-leading orders of a small $c$ expansion of relativistic actions \cite{deBoer:2021jej}. Incidentally the electric kind of actions are said to contain only time derivative, and magnetic ones are described as containing both time and space derivative of fields. However it may be too soon to comment on these distinctions, and we will come back to this discussion elsewhere. 

\section{Discussion and Conclusions }\label{sec8}

In this paper, we set out to specifically understand Carrollian fermions in two dimensions, and we've done so in a thoroughly systematic way throughout. The degenerate nature of Carroll spacetimes allow for degenerate representation of Clifford matrices, and their structures are far more non-trivial than their relativistic cousins. After classifying these representations,  we have clarified the continuous and discrete symmetries associated to Carroll spinors. We then wrote down explicitly Carroll boost invariant actions for these spinors, in line with previous explorations into Null Superstrings. The massless versions of these theories are shown to be invariant under Carroll Conformal symmetries (or BMS symmetries) generated by the conserved charges. We then explored the quantizations of these theories assuming a highest weight representation of the vacuum, which lead to a clear multiplet structure for the Inhomogeneous spinors, and a chiral truncation to a single Virasoro for the homogeneous one. 
\medskip

We also concentrated on the formalism of arriving at Carroll structures under infinitely boosted CFT counterparts. We defined interpolating representations of gamma matrices that flow from relativistic representations to BMS representations as the associated lightcones close upon themselves. Starting from Majorana or Majorana-Weyl representations in two dimensions and following through with this procedure lead us to a web of degenerate Clifford representations. We also wrote down the transformed actions in both representations. At the BMS point, Majorana representations give rise to our Inhomogeneous action, while Majorana-Weyl paves way to a Chiral spinor action.
\medskip

However, this is just a first step into a very intricate regime where Carroll fermions offer a glimpse of exotic physics, and this step mainly helps us to set up our basic structures and tools. Once that is done, there is a bunch of ways we can go ahead and study intriguing systems where Carroll fermions would appear. One immediate extension that comes to mind is to extend these structures to Supersymmetric Carroll actions. Although some work has been done from worldsheet point of view  \cite{Bagchi:2016yyf, Bagchi:2017cte, Bagchi:2018wsn} (see also \cite{Bagchi:2022owq, Hao:2022xhq} for field theoretic discussions), there is still a lot to learn about these theories. For example, quantization and vacuum structures of such theories, even from the string theory point of view, is not clear at all. One would imagine there is a systematic way to classify possible consistent vacua for a Super-BMS theory on the worldsheet, in line with what was done for pure bosonic strings in \cite{Bagchi:2020fpr}, but that discussion hasn't materialized yet. 
After all, we would be interested to establish some version of Flat Holography embedded into Carrollian structures, and it goes without saying that laying the foundations of Carroll Superconformal theories would be invaluable progress along that direction. 
\medskip

Another important directions would be to find Carrollian fermions in `real' systems. The companion paper to this work \cite{4dcarroll} has already made a major and unexpected stride along this direction in terms of finding emergent Carroll symmetries in Flat-band physics, i.e. physics of strongly correlated dispersionless electrons. It can then be expected that generic spin chain or lattice systems in one spatial dimensions would show Carrollian behaviour in some exotic points on the dispersion curve where the Dirac cones are flattened. Flat-band physics in one and higher spatial dimension(s) encompasses various other intriguing phenomena including spin liquids and fractional quantum hall systems. It would be exciting to link these table-top physical phenomenon to emergence of a guiding Carroll symmetry, thereby making it the true ``\textit{One symmetry to rule them all}". We hope to come back to these questions with immediate effect. 
\section*{Acknowledgements}
The authors are indebted to Arjun Bagchi for his insight and inputs at every stage of this work and also for his comments on the manuscript. They would also like to thank Rudranil Basu, Daniel Grumiller, Minhajul Islam and Hisayoshi Muraki for very fruitful discussions. The work of ArB is supported by the Quantum Gravity Unit of the Okinawa Institute of Science and Technology Graduate University (OIST). ArB would like to thank TU Wien for hospitality during the early stages of this project. SD is supported by grant number 09/092(0971)/2017-EMR-I from Council of Scientific and Industrial Research (CSIR), India and the Junior Research Fellowship Programme from ESI Vienna. SD would also like to acknowledge the hospitality of Institute for Theoretical Physics, TU Wien and Erwin Schroedinger International Institute for Mathematics and Physics, Vienna, where a significant part of this work was carried out. SM is supported by grant number 09/092(1039)/2019-EMR-I from Council of Scientific and Industrial Research (CSIR).
\begin{appendix}
\section{Adjoint and Charge conjugation for Homogeneous Spinors}\label{apA}
We discussed earlier that since one of the gamma matrices in the homogeneous case is identically null, fixing the discrete structures for this case is actually troublesome. 
Moreover, the two sub-classes of homogeneous spinors have very different characteristics under Dirac and Charge conjugation operation. Let us discuss their structures case by case in what follows. For $\gb^0=\mathbf{I}, ~\gb^1 = \mathbf{0}$, we have $\gb^{{\mu}\dagger} = + \Lambda\gb^{\mu}\Lambda^{-1}$ and $\gb^{\mu^T} = +\mathcal{C}\gb^{\mu}\mathcal{C}^{-1}$, so that the adjoint matrices and relevant reality conditions become: 
\be{} \Lambda=
			\begin{pmatrix}
				0 & b\\
				c & 0
			\end{pmatrix}~~
			\mathcal{C}=
			\begin{pmatrix}
				0 & 1/b\\
				1/c & 0
			\end{pmatrix}
			\implies  \psi_0^{*}=\psi_{0} ,\quad \psi^*_1=\psi_1.	\ee\\
			On the other hand, for the non-trivial case, $\gb^0=\s_k, ~\gb^1 = \mathbf{0}$, we have $\gb^{{\mu}\dagger} = (-1)^k \Lambda\gb^{\mu}\Lambda^{-1}$ and $\gb^{\mu^T} = -\mathcal{C}\gb^{\mu}\mathcal{C}^{-1}$, so that we have:
\be{} \Lambda=
			\begin{pmatrix}
				0 & b\\
				-b & 0
			\end{pmatrix}~~
			\mathcal{C}=
			\begin{pmatrix}
				0 & 1/b\\
				-1/b & 0
			\end{pmatrix}
			\implies  \psi_0^{*}=\psi_{0} ,\quad \psi^*_1=\psi_1	\ee
In both these cases $b,c \in \mathbf{C}$. These are still fairly general and unconstrained to be fair. But the fact we learn here js when we put $\gb^0=\s_k, ~\gb^1 = \mathbf{0}$ in the homogeneous case, we can use our $\Lambda$ to define the Dirac adjoint, which we have done while writing down the homogeneous action with real spinors in \eqref{actionhom}.

\end{appendix}
\bibliographystyle{JHEP}
\bibliography{NLF}
 
\end{document}